\documentclass[prd,preprintnumbers,superscriptaddress,amsmath,amssymb,nofootinbib,twocolumn,showpacs]{revtex4-2}

\usepackage{dcolumn}
\usepackage{bm}
\usepackage{graphicx}
\usepackage{float}

\usepackage{footmisc}

\usepackage{color}
\usepackage{natbib}
\usepackage{twoopt}
\usepackage{lipsum}

\usepackage[pdftex,
            breaklinks=true,%
            colorlinks=true,%
            linkcolor=red,
            urlcolor=blue,
            citecolor=blue,
            pdfauthor={Facchinetti, Stref, Lavalle},%
            pdftitle={Template for article in XXX}%
           ]{hyperref}

\graphicspath{{./}{Figs/}{figures/}{../figures/}}

\usepackage{shortcutLatex}

\begin{document}

\title{Tidal stripping of dark matter subhalos by baryons from analytical perspectives: disk shocking and encounters with stars}

\author{Ga\'etan Facchinetti}
\email{gaetan.facchinetti@ulb.be}
 \affiliation{Service de Physique Th\'eorique, Universit\'e Libre de Bruxelles, \\ C.P. 225, B-1050 Brussels, Belgium}
\affiliation{Laboratoire Univers \& Particules de Montpellier (LUPM),
  CNRS \& Universit\'e de Montpellier (UMR-5299),
  Place Eug\`ene Bataillon,
  F-34095 Montpellier Cedex 05 --- France}

\author{Martin Stref}
\email{martin.stref@lapth.cnrs.fr}
\affiliation{LAPTh,
  Universit\'e Savoie Mont Blanc \& CNRS,
  Chemin de Bellevue,
  74941 Annecy Cedex --- France}

\author{Julien Lavalle}
\email{lavalle@in2p3.fr}
\affiliation{Laboratoire Univers \& Particules de Montpellier (LUPM),
  CNRS \& Universit\'e de Montpellier (UMR-5299),
  Place Eug\`ene Bataillon,
  F-34095 Montpellier Cedex 05 --- France}


\begin{abstract}
The cold dark matter (CDM) scenario predicts that galactic halos should host a huge amount of subhalos possibly lighter than planets, depending on the nature of dark matter. Predicting their abundance and distribution has important implications for dark matter searches and searches for subhalos themselves, as they could provide a decisive test of the CDM paradigm. A major difficulty in subhalo population model building is to account for the gravitational stripping induced by baryons, which strongly impact on the overall dynamics inside galaxies. In this paper, we focus on these ``baryonic'' tides from analytical perspectives, summarizing previous work on galactic disk shocking, and thoroughly revisiting the impact of individual encounters with stars. For the latter, we go beyond the reference calculation of Gerhard and Fall (1983) to deal with penetrative encounters, and provide new analytical results. Based upon a full statistical analysis of subhalo energy change during multiple stellar encounters possibly occurring during disk crossing, we show that subhalos lighter than $\sim 1$~M$_\odot$ are very efficiently pruned by stellar encounters. This modifies their mass function in a stellar environment. In contrast, disk shocking is more efficient at pruning massive subhalos. In short, if reasonably resilient, subhalos surviving disk crossing have lost all their mass but an inner cuspy part, with a tidal mass function strongly departing from the cosmological one. If fragile, stellar encounters make their number density drop by an additional order of magnitude with respect to disk-shocking effects only (\eg~at the solar position in the Milky Way). Our results can be incorporated to any analytical or numerical subhalo population model, as we show for illustration. This study complements those based on cosmological simulations, which cannot resolve dark matter subhalos on such small scales.
\end{abstract}

\maketitle
\preprint{LUPM:22-001, ULB-TH/22-02}
\section{Introduction}
\label{sec:Intro}
The cold dark matter (CDM) scenario is tied to the theory of hierarchical structure formation \cite{Peebles1982,BlumenthalEtAl1984,PressEtAl1974,BondEtAl1991a,LaceyEtAl1993,MoEtAl2010}, in which small-scale halos, much smaller than typical galaxies, collapse first in the denser and younger universe before larger and larger halos assemble through mergers and accretion. Consequently, the distribution of dark matter (DM) in galactic halos such as that of the Milky Way (MW) is expected to exhibit inhomogeneities in the form of smaller structures spanning a wide range of masses. Although subject to tidal stripping, a significant fraction of these subhalos are to survive and populate their host galaxies in number, as explicitly verified in cosmological simulations down to their numerical resolution limits \cite{GaoEtAl2004,DiemandEtAl2007,SpringelEtAl2008,AnguloEtAl2009}.

Paradoxically enough, the structuring of CDM on small scales may also lead to a mismatch between theoretical predictions and observations, all this being termed as the ``CDM small-scale crisis'' (see \eg~Ref.~\cite{BullockEtAl2017} and references therein). However, if the related core-cusp \cite{deBlok2010} and diversity \cite{OmanEtAl2015} problems are certainly serious issues, especially when contrasted with the impressive regularity observed in some other scaling relations with baryons \cite{LelliEtAl2017}, controversial aspects related to subhalos (counting, etc.) may, as for them, find reasonable explanations in terms of baryonic effects or feedback \cite{ZavalaEtAl2019a}. It is obviously necessary to inspect DM-only solutions to these potential problems (see \eg~Ref.~\cite{HuEtAl2000} or \cite{SpergelEtAl2000}), but it is not less important to improve our understanding and description of CDM physics itself on small scales to prepare for additional tests. In this respect, having reliable predictions of the properties of subhalo populations in galaxies, especially of those subhalos light enough not to host baryons, is of particular interest. Indeed, subhalos can imprint gravitational signatures \cite{VanTilburgEtAl2018,DrorEtAl2019}, boost potential DM annihilation signals \cite{SilkEtAl1993,BergstroemEtAl1999a,LavalleEtAl2008,Ando2009,BuckleyEtAl2010,IshiyamaEtAl2010,PieriEtAl2011,LavalleEtAl2012,BerlinEtAl2014,BartelsEtAl2015,StrefEtAl2017,CaloreEtAl2017,AndoEtAl2019,HuettenEtAl2019,CoronadoBlazquezEtAl2019,FacchinettiEtAl2020}, or turn into intermittent local DM winds \cite{Green2002,IbarraEtAl2019}, hence providing additional ways to test specific realizations of the CDM scenario.

High-resolution cosmological simulations provide very important clues to understand the formation and evolution of subhalos, but are limited by three aspects: (i) finite spatial or mass resolution; (ii) the fact they can hardly be matched to specific real galaxies with strongly constrained kinematic properties and specific histories; (iii) changing the input cosmological parameters or the properties of the primordial power spectrum of density fluctuations is very expensive numerically (see e.g. \citeref{VogelsbergerEtAl2016} for alternatives). The former aspect may have significant impact on the way subhalo properties are inferred from simulations (see \eg~Refs.~\cite{vandenBosch2017,vandenBoschEtAl2018,vandenBoschEtAl2018a}), while the second makes it potentially dangerous to blindly extrapolate simulations' results (for instance subhalo spatial distributions, mass functions, etc.) to specific and constrained objects like the MW \cite{StrefEtAl2017}. Even with improved resolution and advanced empirical implementation of baryonic physics, cosmological simulations will hardly be able to probe most of the substructure mass, which could theoretically reside in subhalos with virial masses as light as $\sim 10^{-12}$ M$_\odot$ \cite{HofmannEtAl2001,GreenEtAl2005,BringmannEtAl2007a,Bringmann2009}.

Deepening our physical understanding of the outcomes of simulations is therefore desirable to consistently interpolate their properties down to smaller scales or onto real objects. In the meantime, it is important to develop alternative though complementary analytical or semi-analytical approaches, since these can deal with scales unresolved by simulations, and are also well suited to study other effects like changes in cosmological parameters, in the primordial power spectrum, etc. These alternative approaches are particularly interesting to investigate the effects of subhalos in DM searches and to conceive related tests of the CDM scenario itself \cite{BerezinskyEtAl2003,vandenBoschEtAl2005a,PenarrubiaEtAl2005,ZentnerEtAl2005,Benson2012,ZavalaEtAl2014,BerezinskyEtAl2014,BartelsEtAl2015,StrefEtAl2017,HiroshimaEtAl2018,VanTilburgEtAl2018,DrorEtAl2019}.

In this paper, we will resort to analytical methods to study those gravitational tides experienced by subhalos and generated by the baryonic components of galaxies, which are expected to strongly affect the subhalo properties within the scale radii of galaxies. This notably concerns regions where DM and/or subhalo searches are currently conducted. We will address two different physical phenomena with two different timescales. First, we will briefly review the pruning of subhalos generated by those tidal shocks triggered by crossings of galactic disks in spiral galaxies, called disk-shocking effects. During such crossings, which may last for rather long times with respect to the deep inner orbital timescale in subhalos, the stars and gas confined into disks act collectively as a smooth gravitational field. The analytical procedure that we present to account for disk shocking was actually developed in a previous work \cite{StrefEtAl2017}, which we slightly refine here. The second phenomenon is technically more involved, and regards the tidal stripping induced by individual encounters of subhalos with stars as they pass by each other. The effects of such encounters on subhalos, which occur on much shorter timescales, have already been considered in both analytical \cite{SchneiderEtAl2010,GreenEtAl2007,BerezinskyEtAl2014} and numerical studies \cite{GoerdtEtAl2007,AngusEtAl2007b,ZhaoEtAl2007,StenDelos2019}, in which they were shown to be significant. Here, we aim at revisiting this physical problem, notably by improving over an earlier reference calculation meant to describe a singular encounter and presented in Ref.~\cite{GerhardEtAl1983} (GF83 henceforth), and widely used in subsequent literature. We further aim at gauging the impact of these baryonic tidal effects on the whole subhalo population of a template galactic host. To do so, we will integrate these new results in the analytical subhalo population model that we designed in a previous work \cite{StrefEtAl2017} (SL17 hereafter), tuned to describe the subhalo population of the MW consistently with kinematically constrained MW mass models comprising both DM and baryons \cite{McMillan2017}. This model can easily be adapted to any host halo object, irrespective of its mass and baryonic content.

The paper is organized as follows. In \citesec{sec:diskshocking} we shortly introduce the very bases of the SL17 model, which are more detailed in \citeapp{app:SL17}, and describe the way disk-shocking effects can be analytically described and accounted for in subhalo models. In \citesec{sec:SingleStar}, we turn to individual stellar encounters, and present the computation of the total energy kick felt by test particles bound to a subhalo that passes by a single star. Then, in \citesec{sec:StellarPop_IntegratedEffect}, we address the computation of the impact parameter distribution and the probability of encounter before evaluating the total energy kick induced by one crossing of the stellar disk. The consequences in terms of the SL17 population model are discussed and illustrated in \citesec{sec:StellarPop_SubhaloModel}, before concluding in \citesec{sec:concl}. Further technical details are given in the appendix sections.

\section{Disk-shocking effects on a subhalo population} 
\label{sec:diskshocking}

In this section, we shortly review the effect of disk shocking, as implemented in the SL17 model. We first summarize the SL17 subhalo population model (more details can be found in \citeapp{app:SL17}), and briefly discuss the gravitational tides generated by the global gravitational potential of the host galaxy (here the MW) before addressing disk shocking. We then propose an easy way to quickly implement these effects, induced by smooth gravitational potentials, into subhalo population models. We illustrate our results with the SL17 model.

\begin{figure}[!t]
\begin{center}
\includegraphics[width=0.49\textwidth]{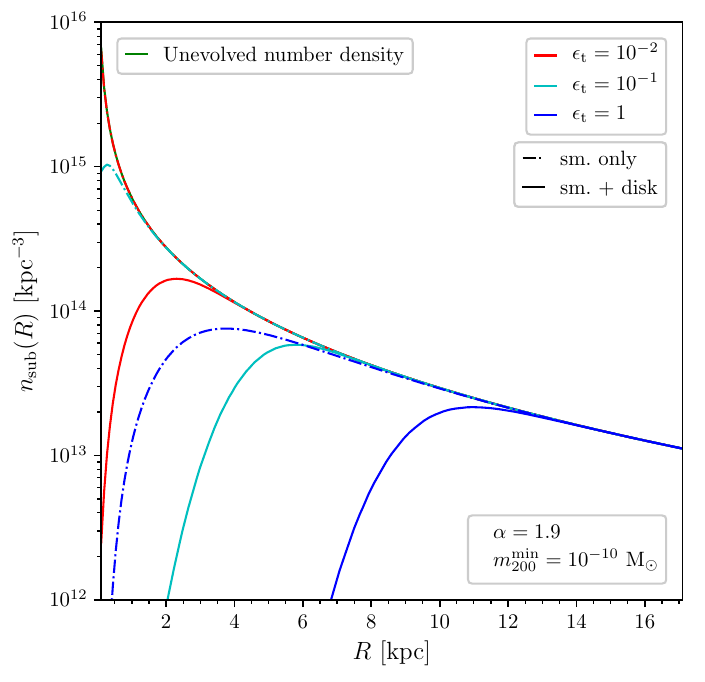}
\caption{\small Number density of subhalos as a function of the galactocentric radius. The host halo is taken as that of the MW, and we assume a power-law subhalo mass function of index $\alpha = 1.9$, with a minimal virial subhalo mass of $m_{200}^{\rm min = }10^{-10}$ M$_\odot$ (see \citeapp{app:SL17}). The initial cosmological distribution (in the hard-sphere approximation) is shown as the green solid curve, which gets first truncated by tidal disruption induced by the smooth gravitational potential of the host halo (dot-dashed curves - ``sm. only''), and then further truncated by the effects of disk shocking (solid curves - ``sm.+disk''). Several tidal disruption efficiencies are considered, from small with $\epsilon_{\rm t}=0.01$, to high with $\epsilon_{\rm t}=1$.}
\label{fig:NumberDensityWithoutStars}
\end{center}
\end{figure} 

The semi-analytical SL17 model \cite{StrefEtAl2017} (see also \citerefs{StrefEtAl2019,HuettenEtAl2019,FacchinettiEtAl2020}) was designed to consistently incorporate a smooth DM halo, baryonic components (disk/s, bulge/s), together with a subhalo population covering the entire mass range allowed by particle DM models, into a global galactic mass model for a target galaxy. It was calibrated for the MW from the mass model constrained on kinematic data by McMillan in \citeref{McMillan2017}, but, by construction, it can be adapted to any host halo object (see examples in \cite{FacchinettiEtAl2022a,LacroixEtAl2022}). The model assumptions are the following: (i) subhalos are building blocks of galactic halos; (ii) if they were hard spheres, they would simply spatially track the overall DM density profile, and retain their initial properties (cosmological mass function, concentration distribution, inner density profiles, etc. --- which can be assessed from the properties of field subhalos); (iii) tidal stripping and mergers are responsible for altering and depleting them. The SL17 model therefore consists in evolving a subhalo population following an overall spherically symmetric DM density profile, starting from initial cosmological properties (virial masses $m=m_{200}$, concentration $c=c_{200}$, position $R$), and then plugging in tidal effects to redistribute the DM stripped away from subhalos to the smooth DM component. The model is statistical in essence, because subhalos are described by probability density functions (PDFs) for their position $R$, mass $m$, and concentration $c$. Tidal stripping is then responsible for spatially-dependent mass losses, which make subhalos with initially the same virial mass $m$ find themselves with different tidal (hence physical) masses $m_{\rm t}(m,c,R)$ depending on both position and concentration. This procedure makes the overall parametric PDF fully intricate and therefore not separable, as tidal masses exhibit strong dependencies on all the other model parameters. Although the SL17 model can be derived for any assumption for the inner density profiles of subhalos, we will assume Navarro-Frenk-White (NFW) \cite{NavarroEtAl1996a} inner profiles in this paper.

On top of tidal stripping, the SL17 model allows for tidal disruption of subhalos based on a rather simple criterion inspired from numerical simulation studies \cite{HayashiEtAl2003}, in which it was shown that subhalos tidally pruned below their initial scale radii, $r_{\rm s}$, would actually be destroyed. By defining $x_{\rm t}$ as the ratio of the subhalo tidal radius $r_{\rm t}(m,c,R)$ to its scale radius $r_{\rm s}$, it is possible to fix a threshold $\epsilon_{\rm t}$ below which a subhalo can be declared destroyed (i.e. if $x_{\rm t}<\epsilon_{\rm t}$). Although early dedicated studies seemed to indicate $\epsilon_{\rm t}\sim 1$ \cite{HayashiEtAl2003}, such a value has been strongly questioned in more recent studies \cite{vandenBosch2017,vandenBoschEtAl2018,vandenBoschEtAl2018a,ErraniEtAl2020,ErraniEtAl2021}, where it was shown that estimates of the tidal disruption efficiency could be significantly biased by numerical effects, and have been likely overestimated in past studies. In short, this means that $\epsilon_{\rm t}$ could actually take much smaller values. For the sake of illustration, we adopt two reference choices in this paper:
\ben
\begin{cases}
  \epsilon_{\rm t}=1\quad &\text{(for {\em fragile} subhalos)}\\
  \epsilon_{\rm t}=0.01\quad &\text{(for {\em resilient} subhalos)}
\end{cases}\,,
\een
keeping in mind that $\epsilon_{\rm t}$ could take even smaller values. As a result of tidal stripping and disruption, the most concentrated subhalos are found to be the most resistant, as naively expected.

By including both tidal stripping itself and such a simple criterion for tidal disruption, the SL17 model is able to quite naturally explain the flattening of the subhalo number density and its further depletion as one gets closer and closer to galactic centers, which is generically observed in cosmological simulations \cite{DiemandEtAl2004,DiemandEtAl2008b,SpringelEtAl2008}. It also explains the radial dependence of the mass-concentration relation \cite{DiemandEtAl2008b,PieriEtAl2011,MolineEtAl2017} as a selection effect. The main output of the SL17 model is the differential number density of subhalos $n_{\rm sub}$ in their host DM halo, which can be expressed either in terms of virial mass $m$, or in terms of tidal mass $m_{\rm t}$, such that
\ben
\begin{split}
& \frac{\dd^2n_{\rm sub}(m,c,R)}{\dd c\,\dd m_{\rm t}} = \\
 & \qquad \int \dd m \,\frac{\dd^2n_{\rm sub}(m, c, R)}{\dd c\,\dd m}\,\delta\( m-m_{\rm t}(m,c,R)\)\,,
\end{split}
\een
where all relevant parameters appear explicitly---see \citeapp{app:SL17} for more details.

Two distinct tidal effects are accounted for in the original SL17 model. The first and simplest effect is the stripping of subhalos by the smooth gravitational potential of the whole host galaxy. This can be modeled by assigning to each subhalo a tidal radius inferred from the following equation,
\ben
r_{\rm t,sm}(m,c,R)= R\,\left(\frac{m(r_{\rm t,sm})}{3\,M(R)\,(1-\frac{1}{3}\frac{\dd \ln M}{\dd \ln R})}\right)^{1/3}\,,
\label{eq:rt_smooth}
\een
where $m(r_{\rm t,sm})$ is the subhalo mass inside $r_{\rm t,sm}\leqslant r_{200}$, and $M(R)$ is the total mass of the host galaxy within radius $R$ (including both the DM and baryons) \cite{BinneyEtAl2008}. This relation, based on the extrapolation of the Roche criterion to diffuse objects, has been shown to nicely correlate with simulation results \cite{TormenEtAl1998,SpringelEtAl2008}.

The second effect, exclusively due to baryons this time, and one of the subjects of this paper, is the gravitational shock induced at each crossing of the disk, called disk shocking. This gravitational shocking is generated by the smooth potential of a galactic disk, inside which gas and stars act collectively. Indeed, test-mass particles (DM particles here) bound inside a subhalo that crosses such a disk experience a kick in kinetic energy due to the ``rapidly'' changing gravitational potential. In order to evaluate this kick, we use the impulsive approximation: the crossing is considered fast enough for particles to be assumed frozen in the frame of the subhalo. When averaged over a subhalo radial shell, the kick in velocity is given by \cite{OstrikerEtAl1972}
\ben
\Delta \vv_{\rm d} = \frac{2\,g_{\rm d}}{\sqrt{3}\,v_z}\, r\, \ev_z  \, .
\label{eq:Delta_vv_d}
\een
This expression depends on $\ev_z$, the unit vector normal to the galactic plane, $v_z\equiv \vv\cdot \ev_z$, the associated subhalo velocity component, and $g_{\rm d}$, the gravitational acceleration due to the potential of the disk. This translates into a kick in kinetic energy per unit of particle mass,
\ben
\Delta E_{\rm d} =\frac{1}{2}\[\(\vv + \Delta \vv_{\rm d}\)^2 - \vv^2 \] = \frac{1}{2} |\Delta \vv_{\rm d}|^2 + \vv  \cdot \Delta \vv_{\rm d} \, ,
\een
with $\vv$ the initial velocity in the subhalo frame. Averaging over an initial isotropic velocity distribution, the second term vanishes, so that we consider only $\Delta E_{\rm d} \sim (\Delta \vv_{\rm d})^2/2$. However, the impulse approximation often breaks down in the case of disk shocking, especially for test particles located in the inner parts of subhalos or in small subhalos---where the typical orbital time is shorter than crossing time. Then, adiabatic invariance has to be accounted for. Indeed, if inner orbital periods are short enough with respect to crossing time, then particles are further protected against stripping by virtue of angular momentum conservation \cite{Weinberg1994,Weinberg1994a}. In the end, according to \citeref{GnedinEtAl1999b}, the energy gain is more reliably evaluated by
\ben
\Delta E_{\rm d} = \frac{1}{2}(\Delta \vv_{\rm d})^2 A_1(\eta_{\rm d}) =   \frac{4\,g_{\rm d}^2}{3\,v_z^2}\,r^2\,A_1(\eta_{\rm d})\,,
\label{eq:Delta_E_d}
\een
where
\ben
A_1(\eta) = (1+\eta^2)^{-3/2}\leqslant 1
\een
is a corrective suppression factor that encodes the effect of adiabatic invariance at first order, hence the subscript. The subhalo adiabatic parameter for disk shocking, $\eta_{\rm d}$, is given by
\ben
\eta_{\rm d}=t_{\rm d}\,\omega\geqslant 0\,,
\een
where $t_{\rm d}\equiv H_{\rm d}/v_{z}$ is the crossing time with $H_{\rm d}$ the thickness of the disk---for applications to the MW, we set $H_{\rm d}=0.9\,\rm kpc$. The orbital frequency of DM particles $\omega$ at radius $r$ inside a subhalo is approximated by
\ben
\omega(r)= \frac{\sigma_{\rm sub}(r)}{r}\,,
\label{eq:orb_frequency}
\een
where $\sigma_{\rm sub}$ is the unidimensional (1D) velocity dispersion in the subhalo evaluated using Jeans' equation\footnote{While those expressions were used in SL17, $v_z$ and $\omega$ are computed slightly differently in this new analysis. Indeed $v_z=v_z(R)$ was fixed to $\sqrt{1/2}$ times the circular velocity at radius $R$, i.e. the 1D dispersion velocity of an isothermal sphere. Here, in order to be consistent with the rest of the study, we take $v_z(R)=\sqrt{1/3}\sigma_v$  as the average of a Maxwell-Boltzmann, isotropic, distribution with an NFW profile for the total Galactic density. This yields $v_z \sim \sqrt{2/\pi}\sigma_c(R)$, where $\sigma_c(R)$ is the velocity dispersion computed from Jeans' equation. Moreover, $\omega$ was evaluated for an isothermal sphere, while here the true profile of the subhalos is used.} \cite{Jeans1902}---see its definition in \citeapp{app:velocity_pdf}. Whenever $\eta_{\rm d}\gg 1$, \ie~adiabatic shielding is efficient, the energy kick is suppressed by the corrective factor $A_1$.\\

The tidal radius is then evaluated recursively. The number of disk crossings $N_{\rm cross}$ is computed with the assumption that subhalo orbits are circular in a steady galaxy. The algorithm starts with $r_{\rm t, 0} = r_{\rm t, \, sm}$ given by \refeq{rt_smooth}, and for every crossing it evaluates a new (maximal) value of $r_{\rm t}$ by requiring that a radial shell receiving an energy kick greater than the gravitational potential of the structure at that radius is removed. More precisely, we make explicit the dependencies on the radius $r$ and on the tidal extension $r_{\rm t}\geqslant r$ of the energy gain function by writing $\Delta E_{\rm d} = \Delta E_{\rm d}(r, r_{\rm t})$. We denote by $\Phi(r, r_{\rm t})$ the gravitational potential
\ben
\Phi(r, r_{\rm t}) = -\int_{r}^{r_{\rm t}}  \dd r' \, \frac{G m(r')}{r'^2}
\een
and the successive maximal tidal radii are evaluated by solving for $r_{{\rm t}, i+1}$ in the equation
\ben
\Delta E_{\rm d} \(r_{{\rm t}, i+1}, r_{{\rm t}, i}\) = \left|\Phi\(r_{{\rm t}, i+1}, r_{{\rm t}, i}\) \right|
\label{eq:recursion}
\een
for all $i \in [1, N_{\rm cross}]$. The tidal radius today is defined by $r_{\rm t} \equiv r_{{\rm t}, N_{\rm cross}}$.

In \refig{NumberDensityWithoutStars}, we report the effect of tidal stripping by considering the complete subhalo population model summarized by \refeq{massFunction} in the appendix. We actually plot the integrated subhalo number density (already integrated over the concentration PDF),
\ben
n_{\rm sub}(R) = \int \dd m_{\rm t} \dd c \der{^2n_{\rm sub}}{c \dd m_{\rm t} } \, ,
\een
with $R$ the distance to the Galactic center. Two configurations are considered, whether disk-shocking effects are switched on or not. For comparison, the cosmological/unevolved number density is also represented (hard-sphere approximation --- solid diverging curve), which merely tracks the host halo profile. For fragile subhalos (disruption efficiency $\epsilon_{\rm t} = 1$), we find a strong suppression due to disk shocking toward the center of the Galaxy (other solid curves) compared with the cosmological distribution and with smooth stripping induced by the overall halo only (dot-dashed curves). The impact is less important with smaller and smaller values of $\epsilon_{\rm t}$.

The next sections are dedicated to evaluating the effects of stellar encounters on the tidal radius and subsequently on the subhalo population. The relative impact of disk shocking and stellar encounters will be compared in \refsec{StellarPop_SubhaloModel}.

\section{Single star encounter}
\label{sec:SingleStar}

\begin{figure}[t!]
  \centering
   \includegraphics[width=\linewidth]{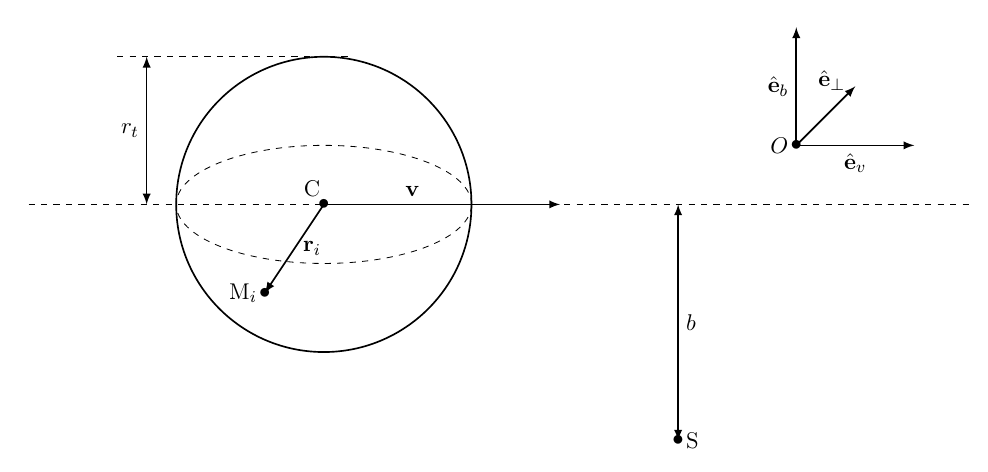}
   \caption{Geometry of the problem. The subhalo in represented as a sphere with center $C$ and radius $r_t$.}
   \label{fig:geometry}
\end{figure}

We focus on the case of a single encounter between a subhalo and an isolated star. The main goal here is to compute the total energy received by every particle in the subhalo during the crossing. We first set up a complete parameterization of the problem before moving on, in a second part, to the full computation and physical analysis of our results. We further compare them with results from previous studies by other authors. From now on, we adopt the convention that for any vector ${\bf  x}$, its norm is given by $x\equiv \lb {\bf x}\rb$.


We start by closely following the original work by Spitzer \cite{Spitzer1958} and its extension by Gerhard and Fall \cite{GerhardEtAl1983} (hereafter GF83), but then extend it even further to penetrative encounters. The geometry we consider for the subhalo-star encounter is summarized in \refig{geometry}. The star $S$ is a point-like object of mass $m_{\star}$ while the subhalo has a radial extension (tidal radius) $r_{\rm t}$, a mass $m_{\rm t}\equiv m(r_{\rm t})$, and its center of mass is located at point $C$. We assume that DM is spherically distributed around $C$ and that spherical symmetry is maintained during the encounter. The center of mass of the entire system defines an inertial frame and we introduce a fixed reference point $O$ in that frame. By Newton's second law, we have for a DM particle $M_i$
\ben
\frac{\dd^2{\bf OM_i}}{\dd t^2} = -\frac{G\,m_{\star}}{\lb {\bf SM_i}\rb^3}\,{\bf SM_i}+\sum\limits_{j\neq i}\frac{G\,m_{\rm p}}{\lb {\bf M_iM_j} \rb ^3}\,{\bf M_iM_j}\,,
\een
where $m_{\rm p}$ is the DM particle mass. The second term on the right-hand side accounts for the self-gravity of the subhalo. In the following, we are going to assume that the typical orbital period of DM particles inside the subhalo is much larger than the duration of the encounter. This implies that the internal dynamics is effectively frozen and that self-gravity can be neglected during the encounter. This is called the \textit{impulse approximation} \cite{Spitzer1958}, and we discuss its validity further in \citesec{ssec:Impulse_approx}. The subhalo dynamics is therefore governed by
\ben
\frac{\dd^2{\bf OC}}{\dd t^2} = \frac{m_{\rm p}}{m_{\rm t}}\left\{-\sum\limits_i\frac{G\,m_{\star}}{\lb {\bf SM_i}\rb^3}\,{\bf SM_i} \right\}\,.
\een 
We introduce the positions of the DM particle and the star with respect to the subhalo center, ${\bf r_i}\equiv {\bf CM_i}$ and ${\bf r_\star}\equiv {\bf CS}$, respectively, and the velocity of a DM particle with respect to $C$,
\ben
{\bf v_i}\equiv \frac{\dd{\bf r_i}}{\dd t}\,,
\een
which obeys
\ben
\begin{split}
\frac{\dd{\bf v_i}}{\dd t} = &  -\frac{G\,m_{\star}}{|{\bf r_i}-{\bf r_\star}|^3}\,\left({\bf r_i}-{\bf r_\star}\right)
\\ & + \frac{\mstar}{m_{\rm t}}\sum\limits_j\frac{G\,m_{\rm p}}{|{\bf r_j}-{\bf r_\star}|^3}\,\left({\bf r_j}-{\bf r_\star}\right)\,.
\end{split}
\een
Taking the continuous limit, the acceleration felt by a test particle at any position ${\bf r}$ in the subhalo can be written as
\ben
\begin{aligned}
\frac{\dd{\bf v}}{\dd t} = & -\frac{G\,m_{\star}}{|{\bf r}-{\bf r_\star}|^3}\,\left({\bf r}-{\bf r_\star}\right) \\
&+ \frac{\mstar}{m_{\rm t}}\int\dd^3{\bf r'}\,\frac{\rho(r')}{|{\bf r'}-{\bf r_\star}|^3}\,\left({\bf r'}-{\bf r_\star}\right)\,,
\end{aligned}
\een
The second term on the right-hand side can be further simplified by using spherical symmetry:
\ben
\frac{\dd{\bf v}}{\dd t} = -\frac{G\,m_{\star}}{|{\bf r}-{\bf r_\star}|^3}\,\left({\bf r}-{\bf r_\star}\right) - \frac{G\,\mstar\,m(r_\star)}{m_{\rm t}}\,\frac{{\bf r_\star}}{r_\star^3}\,,
\een
where $m(r_\star)$ is the subhalo mass contained inside $r_\star$.

The increase in kinetic energy is related to the net change in velocity, which reads
\ben
\delta{\bf v} = \int_{-\infty}^{+\infty}\frac{\dd{\bf v}}{\dd t}\,\dd t\,.
\een
To calculate this integral, we assume that the encounter happens at a high enough speed so that the trajectory of the subhalo can be approximated by a straight line. In that case, we have ${\bf r_\star}(t)=-{\bf b}-{\bf v_r}\,t$, where ${\bf b}$ is the impact vector directed from $S$ to $C$ at the time of closest approach, and ${\bf v_r}$ is the constant relative velocity (see \refig{geometry}). Integration over time leads to 
\ben
\begin{aligned}
\delta{\bf v} = & -\frac{G\,\mstar}{v_r}\frac{{\bf r}+{\bf b}-({\bf r} \cdot \ev_v)\,\ev_v}{r^2-({\bf r} \cdot \ev_v)^2 + b^2+2\,{\bf r} \cdot {\bf b}}\\
&+\frac{2\,G\,\mstar}{v_r}\,\frac{\bf b}{b^2}\,I(b,r_{\rm t})\,.
\end{aligned}
\label{eq:deltav}
\een
In this expression, we have introduced the unitary vector $\ev_v\equiv {\bf v_r}/v_r$. Note that the vector ${\bf r}-({\bf r} \cdot \ev_v)\,\ev_v$ lies in the $(\ev_\perp,\ev_b)$ plane, hence $\delta{\bf v}$, as it should (no component along $\ev_v$, which can easily be understood from symmetry in the time integral). We have also introduced the following integral,
\ben
I(b, r_{\rm t}) \equiv\int_0^\infty\frac{\dd x}{\left(1+x^2\right)^{3/2}}\, \frac{m\left(b\sqrt{1+x^2}\right)}{m(r_{\rm t})}\,.
\een
This integral, which mostly characterizes the relative kick on the subhalo center of mass toward the star, verifies $0\leqslant I(b, r_{\rm t})\leqslant 1$. It is equal to 1 when $b>r_{\rm t}$, and tends to 0 as $b\to 0$.

The result obtained for $\delta{\bf v}$ in \refeq{deltav} can be compared with the work of GF83, in which the authors considered a galaxy perturbed by another galaxy, with both objects modeled in terms of extended Plummer density profiles \cite{Plummer1911}. They also derived an expression for $\delta{\bf v}$ in two limiting cases: $b\ll r_{\rm t}$ (penetrative encounter), and $b\gg r_{\rm t}$ (distant encounter), where $r_{\rm t}$ is the radial extension of the subject galaxy. They proposed an interpolation between these two asymptotic cases to get an expression for any radius $r$. Taking the point-like limit for the perturbing galaxy, their expression becomes identical to ours for both $b\ll r$ and $b\gg r_{\rm t}$. However, our expression for \refeq{deltav} is valid for any values of $b$, $r$ and $r_{\rm t}$, hence no interpolation is required.

\begin{figure*}[!t]
\begin{center}
\includegraphics[width=0.49\textwidth]{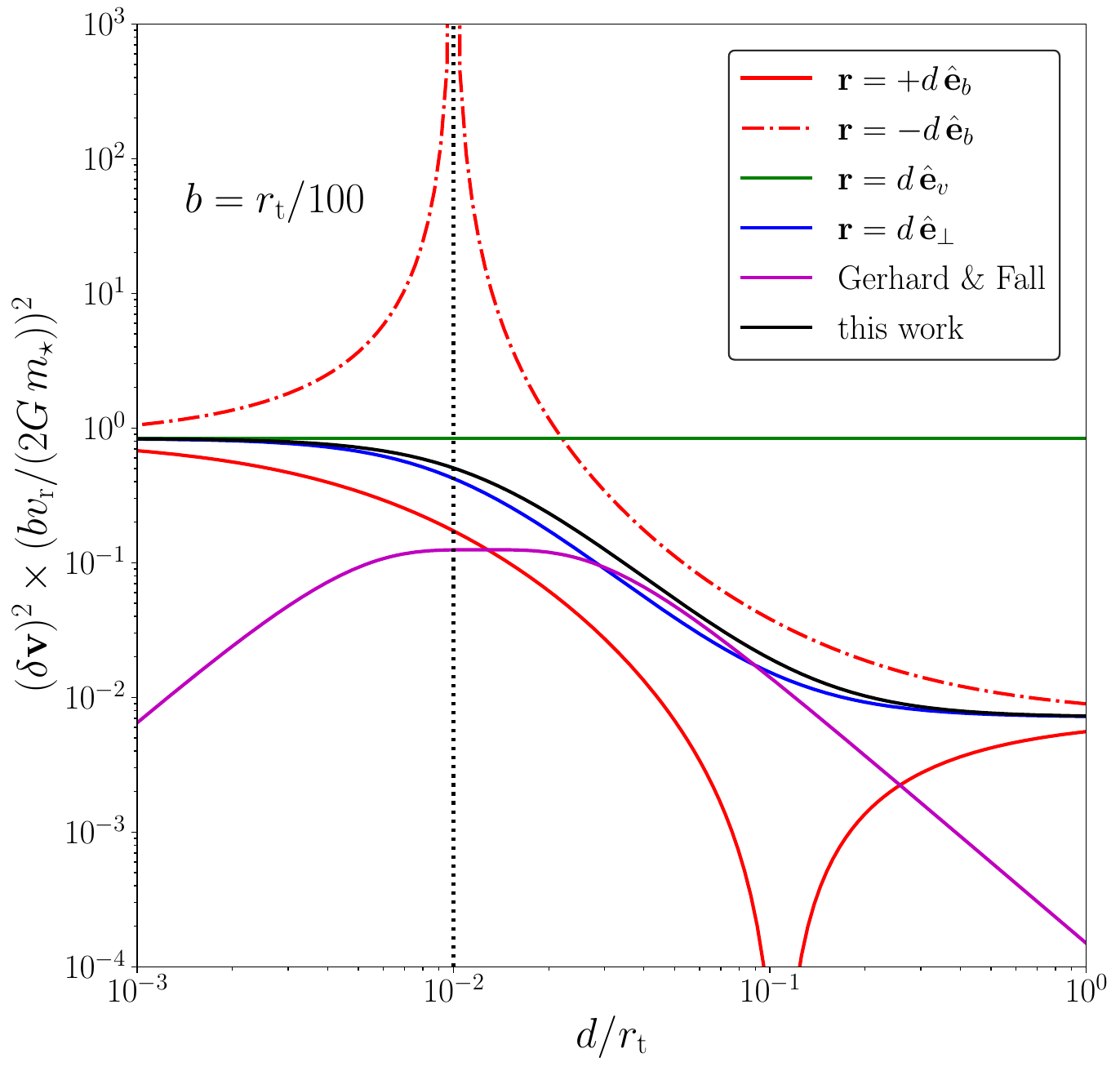}
\includegraphics[width=0.49\textwidth]{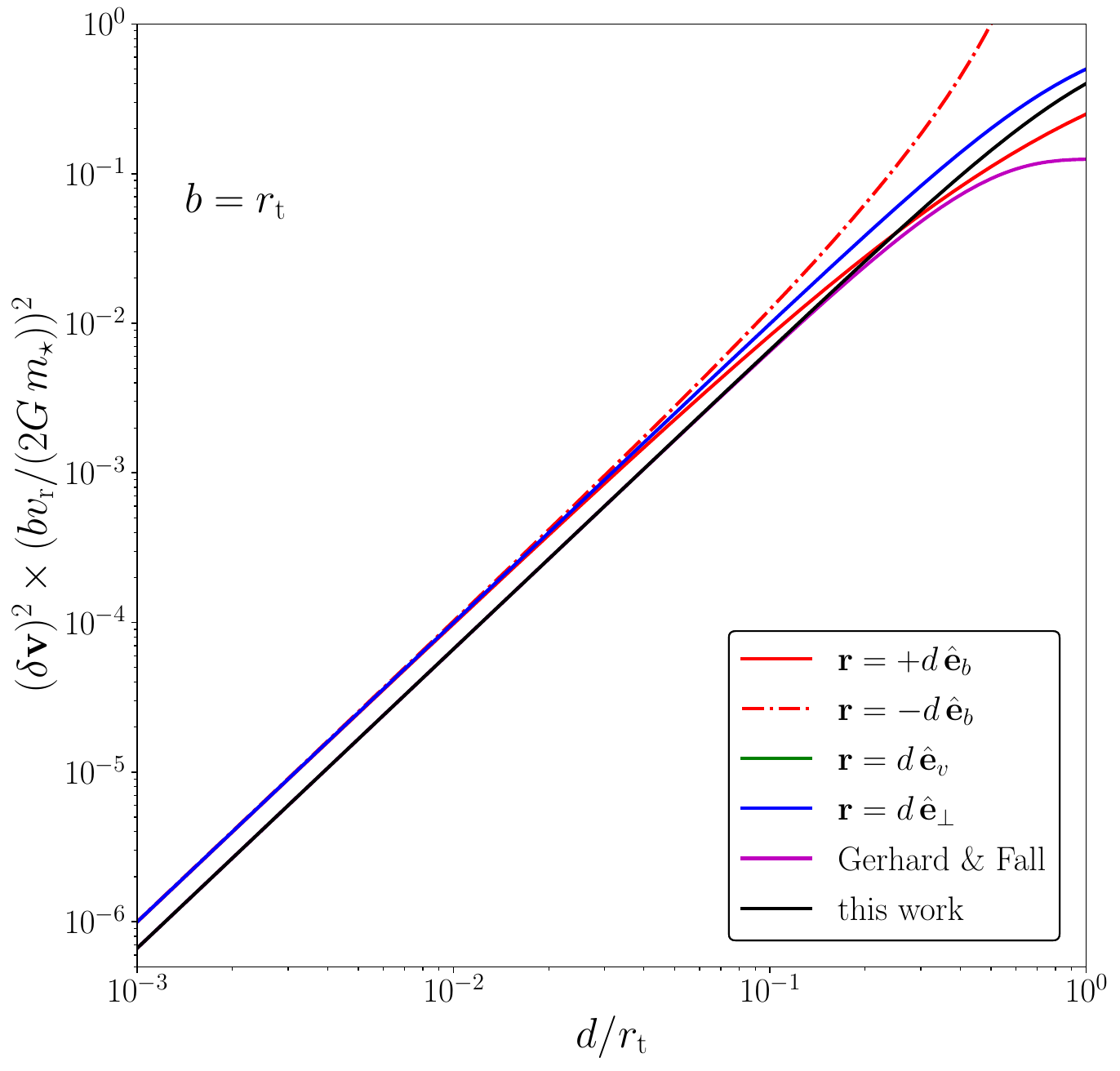}
\caption{\small
$(\delta{\bf v})^2$ along ${\bf b}$ (solid red), in the direction opposite to ${\bf b}$ (dotted-dashed red), along the subhalo trajectory (green) and along $\ev_\perp$ (blue), compared to the prediction by GF83 (magenta) and to the approximation in \refeq{deltav2_approx} (black). Here, we denote by $d$ the distance to the center of the subhalo. These curves only depend on the ratios $r_{\rm t}/r_{\rm s}$ and $b/r_{\rm s}$. We set $r_{\rm t} = r_{\rm s}$ in both panels. The value of the impact parameter, however, is chosen different in the two panels. {\bf Left panel:} Penetrative encounter, $b = r_{\rm t}/100$. {\bf Right panel:} Non-penetrative encounter, $b=r_{\rm t}$. }
\label{fig:deltav2_comparison}
\end{center}
\end{figure*}

Having determined the net change in velocity, we can now access the kinetic energy gain per unit mass,
\ben
\delta E = \frac{1}{2}\left[\left({\bf v}+\delta{\bf v}\right)^2-{\bf v}^2\right] = \frac{1}{2}(\delta{\bf v})^2+{\bf v} \cdot \delta{\bf v} \, .
\label{eq:deltae}
\een
Let us neglect the second term on the right-hand side for the moment (it is expected to average out to 0 for multiple impulsive encounters in a homogeneous field of stars), and focus on the first one. We have
\ben
\label{eq:deltav2_exact}
(\delta{\bf v})^2 &=& \left(\frac{2\,G\,\mstar}{v_{\rm r}\,b}\right)^2\left[I^2+\frac{b^2(1-2\,I)-2\,I\,{\bf r} \cdot {\bf b}}{({\bf r}+{\bf b})^2-({\bf r} \cdot \ev_v)^2}\right]\\
&=& \left(\frac{2\,G\,\mstar}{v_{\rm r}\,b}\right)^2\nn\\
&&\times \left[I^2+\frac{(1-2\,I)-2\,I\,\varepsilon_r \ev_r \cdot \ev_b}{ \varepsilon_r^2 (1- (\ev_r \cdot \ev_v)^2)+2 \varepsilon_r\ev_r\cdot\ev_b+1 }\right]\, ,\nn
\een
where we have introduced the dimensionless parameter $\varepsilon_r\equiv r/b$. This contribution to the energy gain shows that the latter is no longer spherically distributed about the subhalo center, which merely reflects the asymmetry of the interaction. For all particles with positions aligned along the velocity of the subhalo (\ie~$\ev_r=\pm \ev_v$), $(\delta{\bf v})^2$ does obviously not depend on $r$ and is proportional to $(I-1)^2$. In any other direction, we have $(\delta{\bf v})^2\propto (I-1)^2$ when $\varepsilon_r\ll 1$, and $(\delta{\bf v})^2\propto I^2$ when $\varepsilon_r\gg 1$. This is illustrated in \refig{deltav2_comparison} where we show $(\delta{\bf v})^2$ along several directions in cases for which $b\ll r_{\rm  t}$ (penetrative encounter) and $b \gtrsim r_{\rm t}$ (non-penetrative encounter), for an NFW subhalo. Note that, because of our choice for the normalization of the $y$-axis, the plotted curves only depend on $r_{\rm t}/r_{\rm s}$ and $b/r_{\rm s}$, irrespective of the mass and concentration of the subhalo. Here we set $r_{\rm t} = r_{\rm s}$.

The directional dependence of $(\delta{\bf v})^2$ means that spherical symmetry is broken in the kinetic energy distribution after an encounter with a star (assuming an isotropic and spherically symmetric initial velocity distribution). Rather than describing the effect of this phase-space distortion, which goes well beyond the reach of this paper, we derive a spherically-symmetric approximation for $(\delta{\bf v})^2$. Let us first address a classical divergence that characterizes the energy kick of DM particles located along the trajectory of the star in a penetrative encounter, characterized by $r=\pm b/\cos\phi$ in the $(\ev_b,\ev_v)$ plane, which makes the velocity kick of \refeq{deltav} diverge. This is illustrated as the red dot-dashed line in \refig{deltav2_comparison}, which has been taken at ${\bf r}=-{\bf b}$. Another divergence shows up when $b\to 0$, but that one will be discussed later on. Those divergences are just a consequence of considering the star as a point-like object. Should we adopt a more realistic description for our star, with a mass density profile, they would be readily regularized (this could also effectively consist in adding a smoothing length in the gravitational force exerted by the star). The former divergence, of direct interest here, can also be regularized by spherically ``averaging'' our expression. This is done by assuming that all particles on a given radial shell can be described from averaged angular positions. Accordingly, we replace all cosines by their average in the sphere \ie~${\bf r}.{\bf b}\simeq 0$ and $({\bf r}.\ev_v)^2\simeq r^2/3$, a physically relevant trick already used by GF83. This leads to
\ben
(\delta{\bf v})^2 \simeq \left(\frac{2\,G\,\mstar}{v_{\rm r}\,b}\right)^2\left[I^2+\frac{3\,(1-2I)}{3+2\,(r/b)^2}\right]\,.
\label{eq:deltav2_approx}
\een
Under this approximation, $(\delta{\bf v})^2$ is always finite and only depends on $r$. This solution is shown as the black curve in \refig{deltav2_comparison} along with the GF83 solution in magenta. We see that, for $b \ll r_{\rm t}$, our solution reproduces the asymptotic behavior at both large and small radii in almost all directions (particles aligned with the subhalo trajectory receive the same kick irrespective of $r$). On the other hand, the GF83 solution fails to reproduce the correct asymptotic limits in this case. In the opposite case where $b \gtrsim r_{\rm t}$, our solution agrees with GF83.\\

\begin{figure}[!t]
\begin{center}
\includegraphics[width=0.49\textwidth]{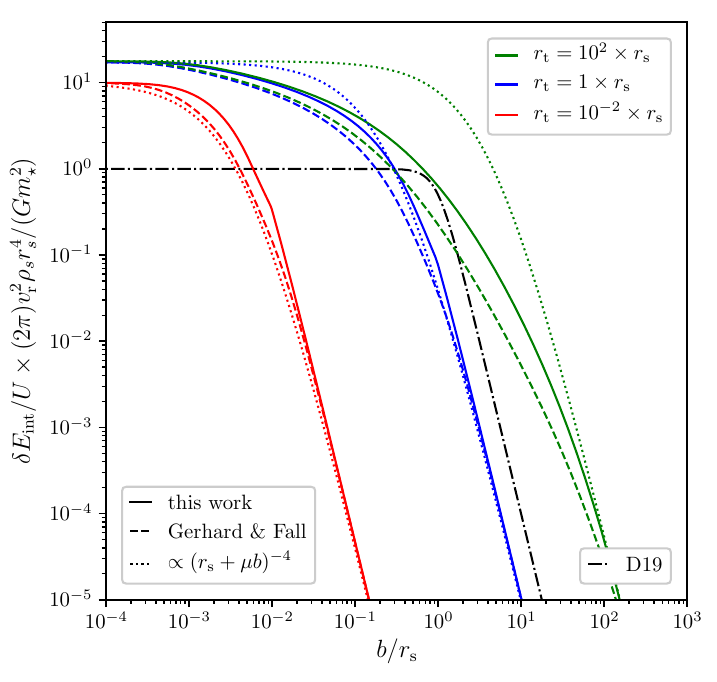}
\caption{\small Ratio of the total kinetic energy integrated on the entire profile over the binding energy for various values of the tidal radius. The  solid curves are obtain with our ansatz  of \refeq{deltav2_approx}, dashed curve are obtained using GF83 result and the dotted lines correspond to a comparison with the usually adopted shape $\propto (\mu b + \rs)^{-4}$. Comparison is made with the characteristic binding energy introduced in \cite{StenDelos2019} (here denoted D19) represented by the dash-dotted line. Due to the normalization of the $y$-axis the only relevant parameter here is the ratio $r_{\rm t}/r_{\rm s}$, for which we consider three different values.}
\label{fig:Egain_over_Ebind}
\end{center}
\end{figure}

It is instructive to compare the total integrated kinetic energy kick of the subhalo with its binding energy, as an indication of its potential disruption or survival to the encounter. We therefore introduce
\ben
\delta E_{\rm int} &=& \frac{1}{2}\int_{\lb{\bf r}\rb\leqslant r_t}\dd^3 {\bf r}\,  \rho(r) (\delta \vv)^2 \\
&=& 2 \pi \int_0^{r_{\rm t}} \dd r\, r^2 \,  \rho(r) (\delta \vv)^2\,,\nn
\een
and the binding energy,
\ben
U &=& \int_{\lb{\bf r}\rb\leqslant r_t}  \dd^3 {\bf r}\, \frac{G_{\rm N}\,m(r)}{r}\, \rho(r)\\
&=& 4\pi\, G_{\rm N} \int_0^{r_{\rm t}}  \dd r\, r\,m(r) \,\rho(r)\, .\nn
\een
The ratio of these two quantities is represented in \refig{Egain_over_Ebind} with respect to the impact parameter -- due to the chosen normalization of the $y$-axis, the only relevant parameter for that figure is $r_{\rm t}/r_{\rm s}$. It scales like $b^{-4}$ when $b \gg r_{\rm s}$ and like a constant (up to a small logarithmic correction) when $b \ll r_{\rm s}$. We actually recover the scaling behavior proposed in \citeref{Moore1993} and used in the context of DM subhalo stripping in \eg~\citerefs{GreenEtAl2007, StenDelos2019} (see also \citeref{GoerdtEtAl2007} for the large $b$ limit):
\ben
\frac{\delta E_{\rm int}}{U} \sim    \frac{G_{\rm N} \mstar^2}{v_{\rm r}^2 \rho_{\rm s}(\mu b+ r_s)^{4}}
\label{eq:DE_over_dU_behav}
\een
with $\mu$ a parameter. Albeit a numerical estimate was given in \citeref{GreenEtAl2007}, $\mu$ is ill-defined for an NFW profile. For example, here we find $\mu = 213$ for $\rt/\rs = 10^{-2}$, $\mu = 3.57$ for $\rt/\rs = 1$ and $\mu = 0.228$ for $\rt/\rs = 10^{2}$.  The dot-dashed curve corresponds to the characteristic binding energy introduced in \citeref{StenDelos2019} (referred to as D19) where the author assumes a slightly different shape $\delta E_{\rm int}\propto 1/(b^4 + \rs^4)$. If the tidal radius is not smaller than the scale radius our solution provides better agreement with \refeq{DE_over_dU_behav} than the GF83 solution.\\

\begin{figure}[!t]
\begin{center}
\includegraphics[width=0.49\textwidth]{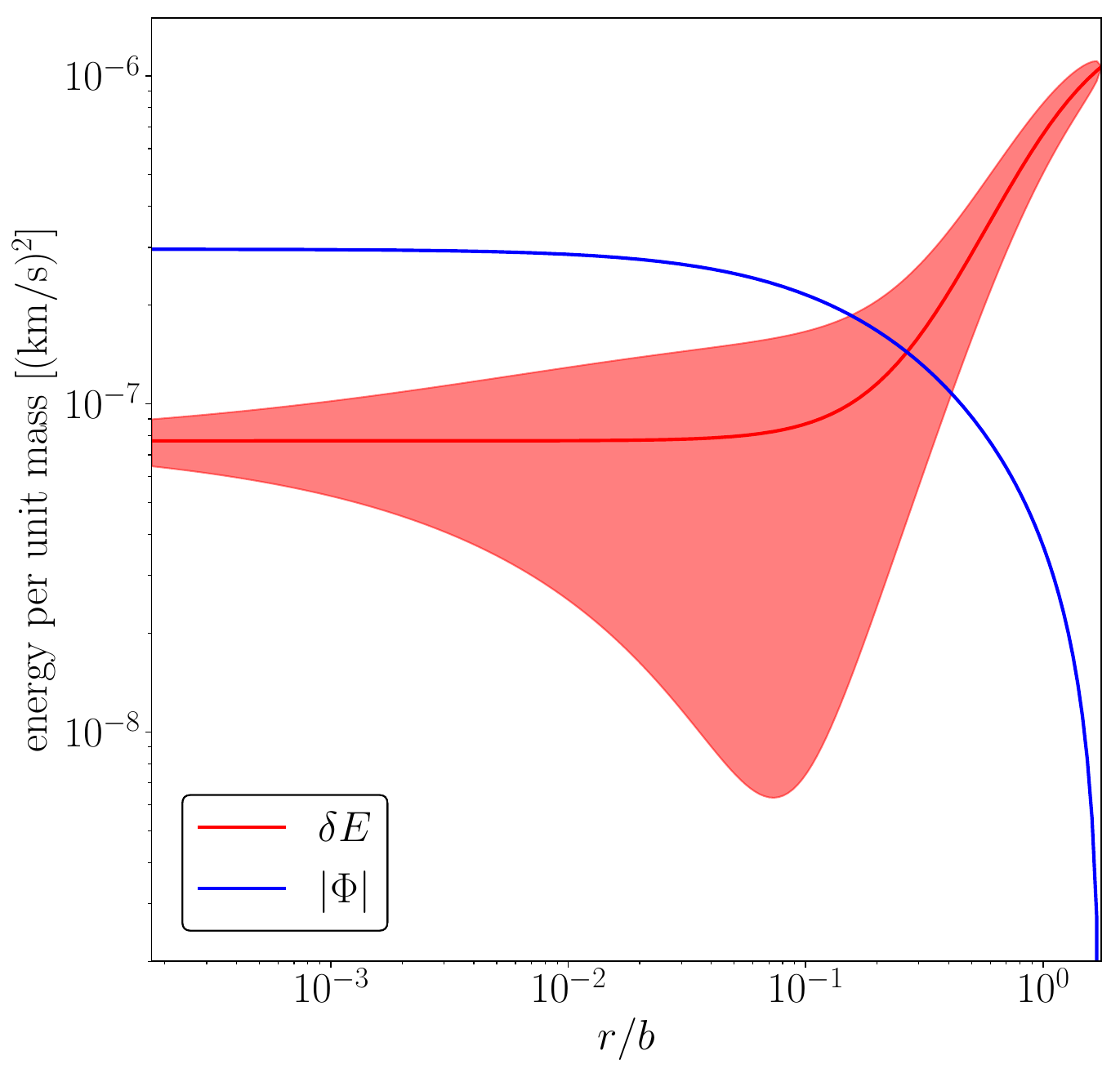}
\caption{\small 
Energy gain (red curve) for a subhalo with parameters $\rho_{\rm s}=5.8\times 10^8\,\rm M_\odot/kpc^3$, $r_{\rm s}=3.5\times 10^{-3}\,\rm pc$ (corresponding to a virial mass $m_{200}=10^{-6}~\rm M_\odot$ and concentration $c_{200} = 60$), $r_{\rm t}=10\,r_{\rm s}$  encountering a star of mass $\mstar=1\,\rm M_\odot$ with an impact parameter $b=2\times 10^{-2}\,\rm pc$ and relative velocity $v_{\rm r}=200\,\rm km/s$. The red-shaded area shows the $1\sigma$ scatter around $\delta E$ due to the inner velocity dispersion. The blue curve shows the gravitational potential defined in Eq.~(\ref{eq:def_gravitational_potential}).}
\label{fig:energy_single_encounter}
\end{center}
\end{figure}

Let us now discuss the second term on the right-hand side of \refeq{deltae}, which is a diffusion term. Since $\delta{\bf v}$ is independent of ${\bf v}$, this term averages to zero over the velocity distribution of DM particles inside the subhalo, however it does contribute to a scatter around $\left<\delta E\right>$. If the velocity distribution at any point in the subhalo is isotropic with a variance $\left< v^2 \right> = 3\,\sigma_{\rm sub}^2$ then we have
\ben
\left<\delta E^2\right>-\left<\delta E\right>^2 = \sigma_{\rm sub}^2\,|\delta{\bf v}|^2\,.
\label{eq:energy_scatter}
\een
The 1D velocity dispersion $\sigma_{\rm sub}$ can be computed from the Jeans equation, see App.~\ref{app:velocity_pdf}. This equation shows that the scatter can be important when $\sigma_{\rm sub}$ exceeds $|\delta{\bf v}|$. The energy gain is compared with the potential energy in \refig{energy_single_encounter} for a subhalo with parameters $\rho_{\rm s}=5.8\times 10^8\,\rm M_\odot/kpc^3$, $r_{\rm s}=3.5\times 10^{-3}\,\rm pc$ (corresponding to a virial mass $m_{200}=10^{-6}~\rm M_\odot$ and concentration $c_{200} = 60$), $r_{\rm t}=10\,r_{\rm s}$ ,  that encounters a star of mass $\mstar=1\,\rm M_\odot$ with an impact parameter $b=2\times 10^{-2}\,\rm pc$ and a relative speed $v_{\rm r}=200\,\rm km/s$. This figure shows that the scatter is important at intermediate radii in the subhalo, while it is small in the central regions and near the edge where the velocity dispersion is small. In the displayed example, we see that a value for the maximal tidal radius after encounter set from $\delta E\sim\lb\Phi\rb$ could change by a rather moderate factor of order ${\cal O}(1)$ due to this scatter in energy.

\section{Effect of the stellar population on a single subhalo}
\label{sec:StellarPop_IntegratedEffect}
This section is devoted to the description of the cumulative effect of multiple stellar encounters as a DM subhalo crosses a galactic disk. We resort to a probabilistic analysis for which we need to understand how a distribution of stars translates into a distribution of impact parameters, relative velocities, and number of encounters. We set up all this in \citesec{ssec:star_pop} before moving to the analysis in terms kinetic energy transfer in \citesec{ssec:total_energy_kick}. We further discuss the validity of the impulse approximation in \citesec{ssec:Impulse_approx}, and compare our final results with those obtained in other studies in \citesec{ssec:comp}. For concrete examples, we will use configurations and parameters relevant for the MW.

\subsection{The stellar population}
\label{ssec:star_pop}
Given a subhalo crossing a stellar galactic disk, we want to know what the distribution of impact parameters for stellar encounters is. Assuming that the disk is an infinite, homogeneous slab with surface mass density $\Sigma_\star$, and that the subhalo moves along a straight line making an angle $\theta$ with respect to the perpendicular to the disk, then the number of encounters between impact parameters $b$ and $b+\dd b$ is
\ben
\dd{\cal N} = \frac{\Sigma_\star}{\overline{\mstar}}\,\frac{2\pi \,b\,\dd b}{\cos(\theta)} \, ,
\label{eq:b_distribution}
\een
where $\overline{\mstar}$ is the mean mass of a disk star. This distribution evidently diverges for $\cos(\theta)=0$, \ie~for orbits contained inside the disk. The infinite and homogeneous assumptions could actually be dropped to get a finite distribution everywhere, but then the final expression would not be analytical and the computation would be much more involved, as shown in \citeapp{app:FullCompDistribImpactParam}. We show that \refeq{b_distribution} is a good approximation as long as $\cos(\theta)\gg b/R_{\rm d}$ where $R_{\rm d}$ is the disk scale radius. This condition is rather generically verified, so we stick to \refeq{b_distribution} in the following.

Our prediction of the kinetic energy gain presented in \citesec{sec:SingleStar} is not valid for arbitrary impact parameters. It is assumed that the encounter is isolated so the impact parameter must be smaller than the typical distance between stars. To compute this distance, we need a model for the considered galactic disk. Here, for concreteness, we use the MW mass model established by McMillan \cite{McMillan2017}. In this model, two exponential stellar disks (thick and thin) are fitted against a number of observational constraints, along with a DM halo, a stellar bulge and two gaseous disks. The best-fit parameters for the stellar disks are given in \citetab{tab:disk_parameters}. Given the axisymmetric mass density profile of stars $\rho_\star(R,z)$, we can define the maximal impact parameter as half of the average interstellar distance,
\ben
\label{eq:bmax}
b_{\rm max}(R) &\equiv& \frac{\int_{-\infty}^{+\infty}\dd z\,\rho_\star\,\frac{1}{2}\,\left(\frac{\rho_\star}{\overline{\mstar}}\right)^{-1/3}}{\int_{-\infty}^{+\infty}\dd z\,\rho_\star} \\
&=& \frac{\overline{\mstar}^{1/3}}{\Sigma_\star(R)}\int_{0}^{+\infty}\rho_\star^{2/3}(R,z)\,\dd z\,.\nn
\een
Using an average mass $\overline{\mstar}\simeq 0.17\,\rm M_\odot$ \cite{Chabrier2003}, we find $b_{\rm max}(8\,\rm kpc)\simeq 1.1\,\rm pc$, which gives a flavor of the local environment.

\begin{figure}[!t]
\begin{center}
\includegraphics[width=0.49\textwidth]{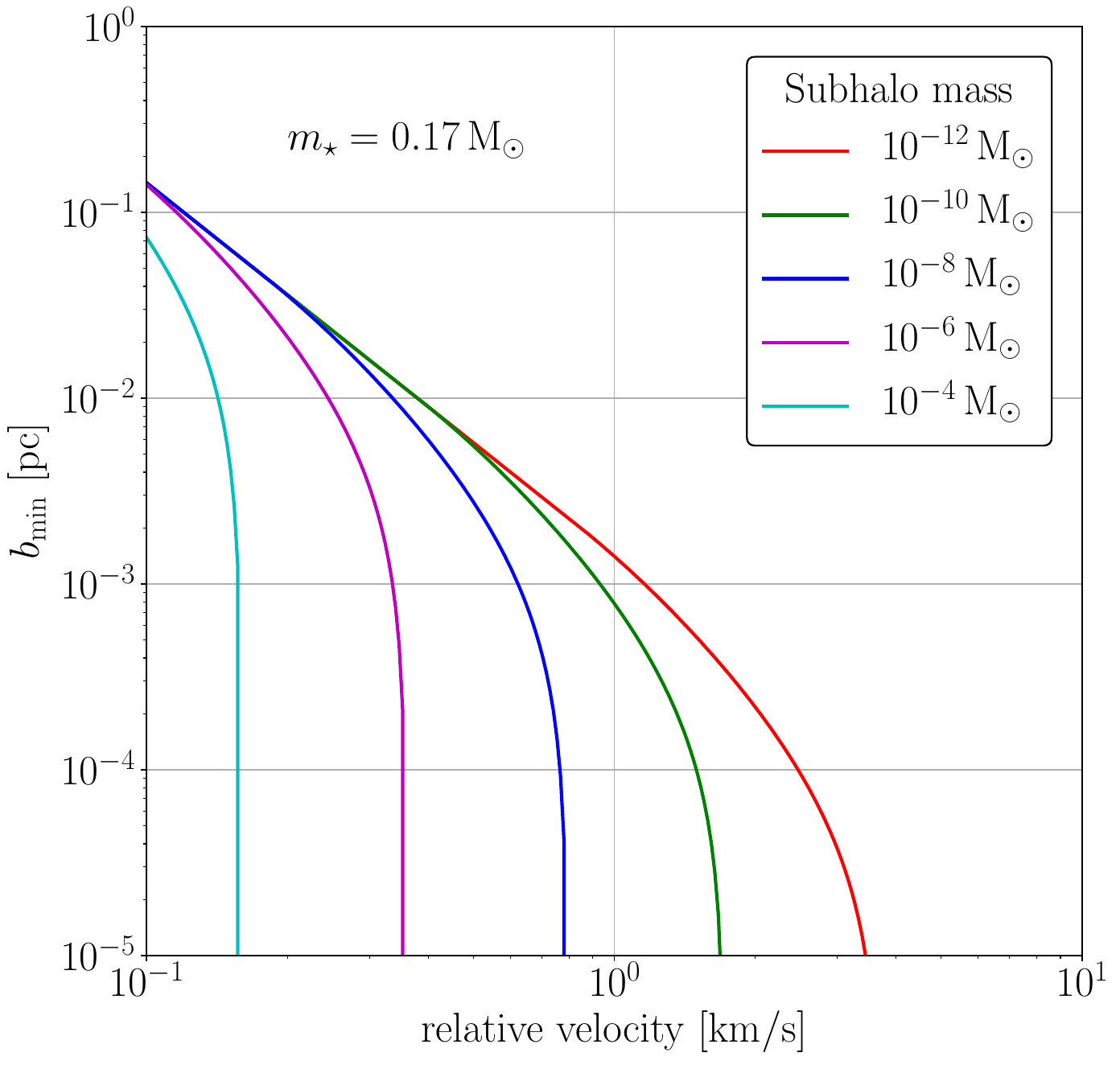}
\caption{\small Minimal impact parameter to avoid significant deflection by the star.}
\label{fig:bmin}
\end{center}
\end{figure}

We also make a straight-line-trajectory assumption when computing $\delta E$. This is reasonable only if the kinetic energy in the center-of-mass frame is much larger than the potential energy, $T\gg|W|$, with
\ben
T = \frac{1}{2}\,\frac{\mstar\,m_{\rm t}}{\mstar+m_{\rm t}}\,v_{\rm r}^2\,,
\een
and 
\ben
W = -G\,\mstar\left[\frac{m(r_\star)}{r_\star}+\Theta(r_{\rm t}-r_\star)\int_{r_\star}^{r_{\rm t}}\frac{\rho(r)}{r}\,\dd^3{\bf r}\right] \, .
\een
$W$ is minimal when $r_\star=b$, where $r_\star$ is the radius of the star. Thus the condition $T\gg|W|$ defines a minimal impact parameter $b_{\rm min}$ below which the subhalo would actually be gravitationally captured. This parameter is shown for several subhalo masses in \refig{bmin}. We see that $b_{\rm min}$ is much smaller than $b_{\rm max}$ unless the relative velocity is smaller than $0.1\,\rm km/s$. Since the typical velocity of subhalos in MW-like galaxies is of order $100\,\rm km/s$, then $b_{\rm min}\sim 0$ in most cases and the assumption of straight-line trajectory is verified. The total number of encounters is 
\ben
{\cal N}=\frac{\Sigma_\star}{\overline{\mstar}}\,\frac{\pi}{\cos(\theta)}\left(b_{\rm max}^2-b_{\rm min}^2\right)\,.
\label{eq:Nencounters}
\een
At $8\,\rm kpc$ in the MW, we find ${\cal N}\simeq 2346\times(0.5/\cos(\theta))$, which is quite significant. 

We also need to specify the relative velocity between stars and subhalos. We assume the subhalo velocity distribution to be an isotropic Maxwell-Boltzmann distribution,  with a dispersion velocity $\sigma(R)$ that can be computed using Jeans' equation, also accounting for the baryonic contribution to the potential---see \citeapp{app:velocity_pdf}. Moreover, assuming that stars follow circular trajectories at a velocity $v_\star(R)$, we get the relative speed distribution
\ben
f(v_{\rm r}) = \sqrt{\frac{2}{\pi}}\,\frac{v_{\rm r}}{\sigma\,v_\star}\,{\rm sinh}\left(\frac{v_{\rm r}\,v_\star}{\sigma^2}\right)\,e^{-(v_\star^2+v_{\rm r}^2)/(2\,\sigma^2)}\,,
\label{eq:relative_speed_distribution}
\een
and the mean relative speed
\ben
\overline{v_{\rm r}} = \sigma\,\sqrt{\frac{2}{\pi}}\left\{e^{-X^2}+\frac{\sqrt{\pi}}{2}(1+2\,X^2)\,\frac{{\rm erf}(X)}{X}\right\}\,,
\label{eq:mean_relative_speed}
\een
where $X=v_\star/(\sqrt{2}\,\sigma)$. At 8 kpc in the MW, the relative speed is $\overline{v_{\rm r}}\simeq 334\,\rm km/s$. The last ingredient is the stellar mass PDF, $p_{\mstar}$,  used to compute $\overline{\mstar}$, which we take from \citeref{Chabrier2003a}. We are now equipped to define the total energy kick received by a subhalo when it crosses the entire stellar disk.

\subsection{Total energy kick and scatter}
\label{ssec:total_energy_kick}

\begin{figure*}[!t]
\centering
\includegraphics[width = 0.47\textwidth]{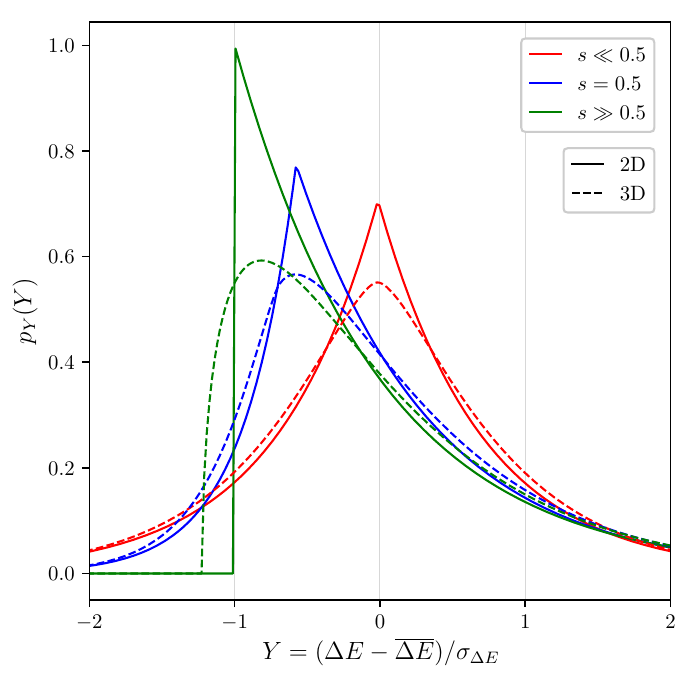}
\includegraphics[width = 0.49\textwidth]{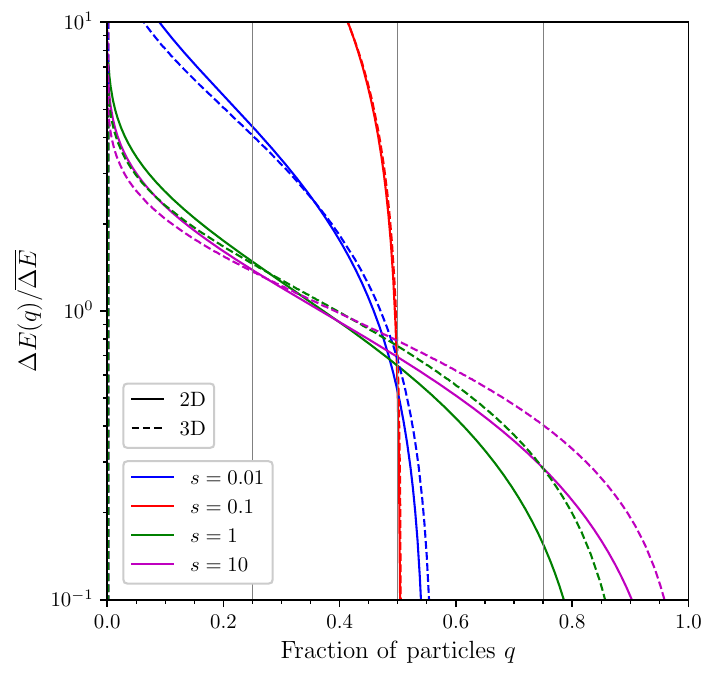}
\caption{\small In both panels, the solid curves rely on $\Delta \vv$ calculated with a 2D isotropic random walk in velocity space.  For comparison, the dashed curves correspond to the case of a 3D isotropic random walk -- see App.~\ref{app:E_and_v_pdfs_for_encounters} for more details. {\bf Left panel}: Probability density function of the centered reduced total energy kick for different values of the parameter $s$. We denoted by $\sigma_{\Delta E}$ the standard deviation of $\Delta E$.  {\bf Right panel}: $\Delta E(q)$ for different values of $s$.}
\label{fig:PDF_InverseCCDF}
\end{figure*}

When crossing the stellar disk, a subhalo encounters $\Nc$ stars, each with a different impact parameter. Thus, a particle inside the subhalo receives a series of velocity kicks $\{\delta \vv_i\}_{1 \le i  \le \Nc}$. We assume that the subhalo does not have time to relax between encounters (impulse approximation), hence the total velocity kick is given by
\ben
\Delta \vv = \sum_{i=1}^{\Nc} \delta \vv_i \, .
\label{eq:sum_dv}
\een
Similarly to \refeq{deltae}, the associated total  energy kick per units of mass is given by
\ben
\Delta E = \frac{1}{2}|\Delta \vv|^2 + \vv \cdot \Delta \vv \, .
\label{eq:Deltae}
\een 
Because encounters are characterized by the statistical distribution of impact parameters and stellar masses, all vectors $\delta \vv_i$ and $\Delta \vv$ are random variables. In the following, we first show how we can estimate a PDF for $\Delta \vv$. The sequence $\{\delta \vv_i\}_{1 \le i  \le \Nc}$ behaves as a random walk in \emph{velocity space}. \refeq{deltav} shows that every $\delta \vv_i$ is confined in the same plane, perpendicular to the relative velocity vector. The random walk is thus two-dimensional\footnote{As stars have their own velocity, the relative velocity vector may vary significantly from one encounter to another if the velocity of the subhalo is not high enough. Therefore the perpendicular plane may fluctuate in orientation and the random walk is not strictly 2D. However, for simplicity, and because it does not impact our results by orders of magnitude (as seen in \refig{PDF_InverseCCDF}), we stick to the 2D hypothesis.} (2D). When the number of encounters $\Nc$ is large enough, this random walk can be described as a Brownian motion as a consequence of the central limit (CL) theorem---we shall check the validity of the CL theorem later on. Since the random walk is isotropic in the plane, the PDF of $\Delta \vv$  can be written
\ben
p_{\Delta \vv}(\Delta \vv) \simeq \frac{1}{\pi \Nc \overline{\delta v^2} } e^{-\frac{|\Delta \vv|^2}{\Nc \overline{\delta v^2}}}\,,
\label{eq:CL_velocity}
\een
and only depends on the second moment of $\delta \vv$, 
\ben
 \overline{\delta v^2} =   \int \dd \mstar  \, p_{\mstar}(\mstar) \int_{b_{\rm min}}^{b_{\rm max}} \dd b \, p_b(b) (\delta \vv)^2 \, ,
 \label{eq:delta_v2_average}
\een
with $(\delta \vv)^2$ given in \refeq{deltav2_approx}, $p_b(b) =( \dd \Nc / \dd b)/\Nc$ [see \refeq{b_distribution}], and $p_{\mstar}(\mstar)$ taken from \citeref{Chabrier2003}. Note that to speed up the numerical calculations, we can approximate $p_{\mstar} (\mstar) \sim \delta_{\rm D}(\mstar- \overline{\mstar})$ in most cases. A first straightforward estimate of $\Delta E$ is obtained by considering the average value of $\Delta \vv$, which yields
\ben
\overline{\Delta E}  \sim  \frac{1}{2} \overline{|\Delta \vv|^2} =  \frac{1}{2} \Nc  \overline{\delta v^2} \, .
\een

\begin{figure*}[!t]
\centering
\includegraphics[width = 0.9\textwidth]{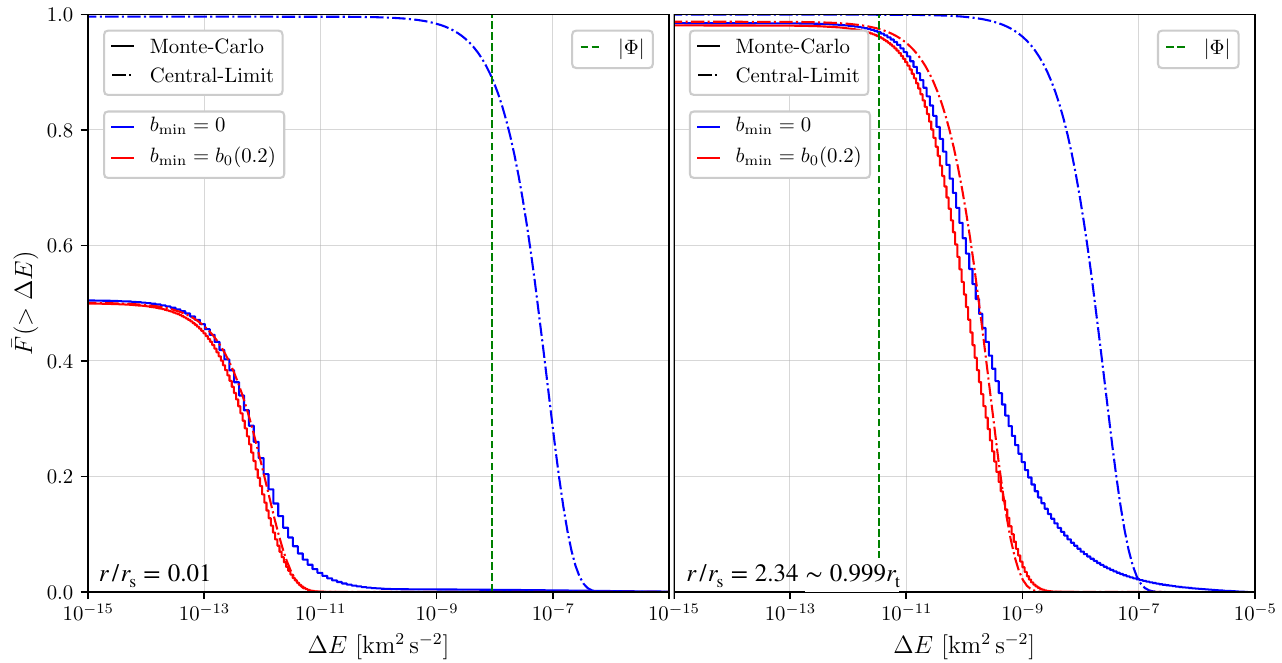}
\caption{\small  Complementary cumulative distribution function of the energy kick received after one crossing of the disk due to the encounter with stars $\Delta E$. In blue is shown the true value (solid) and the approximation using the CL theorem (dashed-dotted) for $b_{\rm min} \sim 0$. In red are similar curve imposing a lower cut-off on the distribution of impact parameters $b_{\rm min} \to b_0(Q=0.2)$. The vertical green dashed line is the value of the gravitational potential $|\Phi|$. {\bf Left panel:} In the inner part of the subhalo $r/r_{\rm s} = 0.01$  {\bf Right panel}:  In the outskirts of the subhalo $r/r_{\rm s} = 2.34$.} 
\label{fig:MC_results}
\end{figure*}

\begin{figure*}[!t]
\centering
\includegraphics[width = 0.49\textwidth]{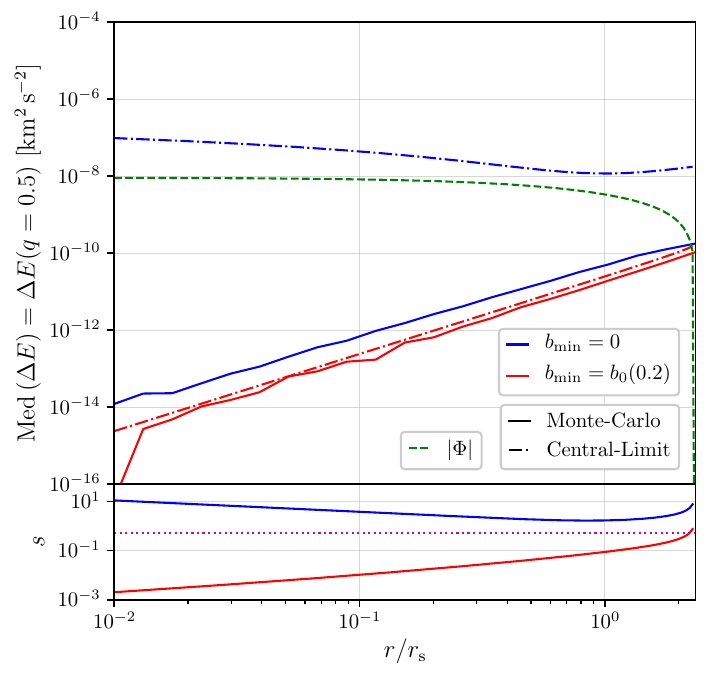}
\includegraphics[width = 0.49\textwidth]{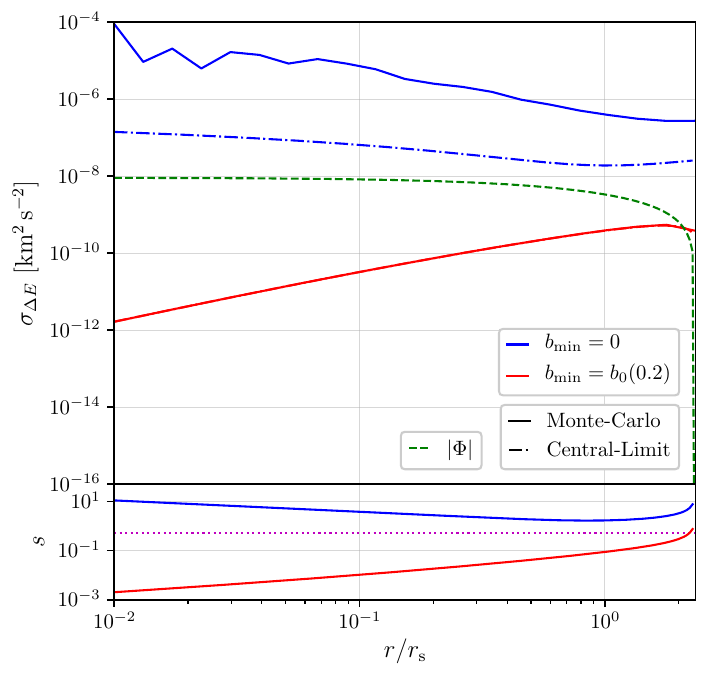}
\caption{\small  {\bf Left panel}. Median energy kick at radius $r$. {\bf Upper panel:}  The true value (solid) and  the CL estimate (long-sort dashed) for $b_{\rm min} \sim 0$ are shown in blue. The corresponding red curves are obtained by imposing a lower cut-off on the distribution of impact parameters $b_{\rm min} \to b_0(Q=0.2)$. The green dashed curve shows the gravitational potential of the subhalo. {\bf Lower panel:} Value of the parameter $s$ in the two configurations with (blue) and without (red) the cut-off on the impact parameters. The magenta dotted curve is the limit $s = 0.5$.  {\bf Right panel.} Same figure for the energy kick standard deviation. Note that the dash-dotted and solid red curves are indistinguishable (for the dispersion, we observe good convergence of the CL estimate when the impact parameter range is truncated).} 
\label{fig:MC_results_vs_r}
\end{figure*}

Now, we have to inspect the impact of the second term in \refeq{Deltae} more closely. In order to take it into account properly, we can actually derive the full PDF of $\Delta E$. The initial velocity distribution being approximated as a Maxwell-Boltzmann distribution of dispersion $\sigma_{\rm sub}$ --- see \citeapp{app:pdf_ccdf_DeltaE}, then
\ben
p_{\Delta E}\(\Delta E \) = 
\frac{\exp\(\frac{\Delta E}{2  \sigma_{\rm sub}^2}- \frac{|\Delta E|} {2  \sigma_{\rm sub}^2} \frac{\sqrt{1+s^2}}{s}\)}{4  \sigma_{\rm sub}^2  s \sqrt{1+s^2}}\,,
\label{eq:pdf_Deltae}
\een
where $s^2 \equiv \Nc \overline{\delta v^2}/(8\sigma_{\rm sub}^2) = \overline{\Delta E}/(4\sigma_{\rm sub}^2)$ is a normalized ratio of the variance of $|\Delta \vv|$ to the variance of the initial velocity $|\vv|$. The associated scatter is
\ben
\sigma_{\Delta E} = \overline{\Delta E}\,\sqrt{1+\frac{1}{2\,s^2}}\,.
\een
This distribution is plotted in the left panel of \refig{PDF_InverseCCDF} in terms of the associated centered reduced variable $Y \equiv (\Delta E - \overline{\Delta E})/ \sigma_{\Delta E}$. When $\sigma_{\rm sub}$ is large (\ie~$\sigma_{\rm sub}^2\gg\overline{\Delta E}$, equivalently $s \ll 1/2$), which is characteristic of big massive subhalos, the dominant source of energy dispersion comes from that in the initial velocity, which is symmetric by virtue of the assumed Maxwell-Boltzmann distribution. When $\sigma_{\rm sub}$ is small ($s \gg 1/2$), however, which is a characteristic of small subhalos, the energy dispersion originates in stellar encounters. The PDF of $Y$ is then peaked toward negative values, and energies below the average are more likely.

To better account for the dispersion in energy kick and for the shift in the distribution, we should evaluate a new density profile for the subhalo by removing the particles kicked out from the gravitational potential (\ie~kicked to speeds larger than the escape speed) -- this possibility is discussed in \citeapp{app:ImpactOnProfile}. However, this procedure requires extensive numerical resources to evaluate the impact of one disk crossing on a single subhalo. Consequently, it is too expensive for the semi-analytical SL17 framework, which deals with a full population of subhalos (typically $\sim 10^{20}$ objects in a MW-like galaxy). In the following, we adopt a simpler criterion and define a {\em unique} energy-kick value per shell as an estimate of the energy kick felt by all particles in that shell.

We introduce the threshold-energy function $\Delta E(q)$. It is defined as the minimal energy kick received by a fraction $q\leqslant 1$ of the particle population located in a given shell -- a full expression is given in \citeapp{app:pdf_ccdf_DeltaE} for an initial Maxwellian velocity distribution. $\Delta E(q)$ is obviously a decreasing function of $q$. The median energy gain is ${\rm Med}(\Delta E) \equiv \Delta E(q=0.5)$.  The ratio $\Delta E(q)/\overline{\Delta E}$ is plotted in the right panel of \refig{PDF_InverseCCDF} for different values of $s$. If $s$ is large enough ($s \gtrsim 0.5$), \ie~if the effect of the stellar encounters is relevant, then $\Delta E(0.25< q < 0.75)$ and $\overline{\Delta E}$ are always close: $\Delta E (q)/ \Delta E \in [0.1, 2]$. In addition, we can show that for any value of $s$,
\ben
\frac{1}{2} < \frac{{\rm Med}(\Delta E)}{\overline{\Delta E}} < \ln(2)\simeq 0.69 \,,
\label{eq:RelationAverageMedian}
\een
so the mean and median values are similar. Therefore, it is both physically meaningful and convenient to define a {\em unique} energy kick for all particles at a given distance from the subhalo center as ${\rm Med}(\Delta E)$. Hence, we define the threshold as
\ben
\Delta E  \equiv \Delta E(q=0.5) \simeq \left\{\kappa=0.7\right\}\, \overline{\Delta E}\,.
\een
This threshold will help us characterize how subhalos are modified.\\

So far, we have assumed that the number of encountered stars per crossing, $\Nc$, is large enough for the CL theorem to apply. As a matter of fact, depending on the position inside a subhalo, this is actually not necessarily the case. Indeed, when $b_{\rm min}$ is close to 0, because the velocity kick satisfies $(\delta \vv)^2 \propto 1/b^4$ for the innermost particles, it gets tremendously large (even when considering a cutoff at the stellar radius). In the meantime, because $p_b(b) \propto b$, for the majority of the disk crossings, a total of $\Nc \sim 10^2 - 10^5$ encounters is not sufficient for rare events, with $b \sim b_{\rm min} \ll b_{\rm max}$, to happen (said differently, the distribution $p_b$ is not fully sampled with such a number of encounters). Therefore, used blindly, the CL-theorem-inferred distribution can overestimate the energy kick felt by particles in the innermost part of the subhalo.

To illustrate and quantify this effect, let us consider a striking example that involves a small subhalo. We consider a typical NFW subhalo before crossing the MW stellar disk, with a typical mass $m_{\rm t} = 1.6 \times 10^{-9} $ M$_\odot$, scale radius $r_{\rm s} = 7.1 \times 10^{-7}$ kpc (cosmological mass $m_{200} = 10^8~\rm M_\odot$ and concentration $c_{200} = 64$), and tidal radius $r_{\rm t} = 2.34192 \times r_{\rm s}$ at a distance $R = 8$ kpc from the Galactic center. These values are typical of a subhalo that would have only been smoothly pruned by the overall potential of the MW. Its velocity relative to the stars in the disk is taken as the average value $\overline{v_{\rm r}}(R = 8 \,{\rm kpc}) = 334$~km/s, and its inclination is fixed to $\cos \theta = 0.5$. Using \refeq{Nencounters}, this implies a number of stellar encounters of $\Nc \sim 2346$. Our goal is to determine the true PDF of $\Delta E$, and compare it with \refeq{pdf_Deltae}. Even though the number of encounters is too small for the PDF to converge to the CL distribution, it is still too high for a full analytical computation. Indeed, this would require to perform $\Nc$ convolutions of the PDF of $\delta \vv$, which is not even possible numerically. A Monte-Carlo (MC) simulation is much better suited, with a total of $5 \times 10^5$ draws to achieve convergence to the true PDF of $\Delta E$. In \citefig{fig:MC_results}, we show the complementary cumulative distribution function (CCDF) of $\Delta E$ (more convenient to display than the PDF, while containing the same information):
\ben
\overline{F}(\geqslant\Delta E) \equiv \int_{\Delta E}^{+\infty}p_{\Delta E}(\Delta E')\,\dd(\Delta E')\,.
\een
It is reported for two radii, one in the inner part of the subhalo with $r = 10^{-2} \times r_{\rm s}$, and another one in the outskirts with $r = 2.34 \times r_{\rm s} \sim 0.999 r_{\rm t}$. The solid curves show the MC results while the dot-dashed ones are the CL-theorem-inferred distributions. For the innermost particles, the CL distribution is shifted toward much higher values of $\Delta E$ with respect to the true distribution. In \refig{MC_results_vs_r}, we report the median energy kick and its dispersion as functions of the distance from the subhalo center. For the same subhalo, the left panel clearly shows that the expectation of the CL-inferred estimate for ${\rm Med}(\Delta E)$  (dashed blue curve) overshoots the gravitational potential over the entire range of radii. The true value, in solid blue, only crosses it toward the outskirts. Therefore, this points out the big error made when naively using the CL-inferred estimate.

Unfortunately, even though an MC simulation can formally be set up to determine the actual PDF, it is far too greedy in terms of computational time to be used for a full subhalo population. Since only encounters with small impact parameters are responsible for the convergence issue, while they have only very small chances to occur, a way out consists in truncating the impact-parameter range from below. If we denote $b_0$ the minimal impact parameter during a single crossing, then we can define the associated PDF from that of the impact parameter given \citeeq{eq:b_distribution}, 
\ben
p_{b_0}(b_0) = \frac{2\Nc}{(1-\beta^2)^\Nc} \frac{b_0}{b^2_{\rm max}} \[1-\(\frac{b_0}{b_{\rm max}}\)^2\]^{\Nc -1}\,,
\een
where $\beta \equiv b_{\rm min}/b_{\rm max} \ll 1$. Similarly to the energy threshold discussed above, we can further introduce the threshold function $b_0(Q)$ defined such that a for ${\cal N}$ disk crossings, a fraction $Q$ of the encounters has impact parameters lower than $b_0(Q)$. This impact-parameter threshold function is given by
\ben
b_0(Q) = b_{\rm max}\[1-(1-\beta^2)(1-Q)^\frac{1}{\Nc}\]^\frac{1}{2}\, .
\label{eq:b0}
\een
A way to converge to a CL-inferred distribution is then simply to enforce a new minimal impact parameter in the expression of $\overline{ \delta v^2}$ given in \refeq{delta_v2_average}. Therefore, we replace $b_{\rm min}$ by $b_0(Q \sim 0.2)$, where the value of $Q = 0.2$ is our recommended tuned value for the CL-inferred approximation to best-match with the correct results. When $b_{\rm min} = 0$, we can show that $b_0(0.2) \sim 1.6 \times  b_{\rm max}/ \Nc$, which scales like the inverse of the number of encountered stars, and makes statistical sense. Thus, we can safely define a corrected CL-inferred distribution whose precision can be appreciated in \refig{MC_results} by comparing (i) the red and blue solid curves (showing the MC results using $b_{\rm min}=b_0$ and $b_{\rm min}=0$, respectively, which should converge to validate the method itself); (ii) the red dashed curves (corrected CL-inferred results) with the blue solid curves (MC results). Convergence is not perfect but the corrected CL-inferred estimate is significantly closer to the true CCDF (solid blue curves) than the nominal one (dot-dashed blue curves). In the right panel of \refig{MC_results_vs_r}, the dispersion is not recovered from the trick itself (compare the blue and red solid MC curves), and the much higher dispersion of the true energy kick (blue solid) is consequently not captured by the corrected CL-inferred distribution (dot-dashed red). This large true dispersion comes from extremely rare events that enhance the mean of the true distribution and shift it away from the median. However, in the left panel, the corrected CL-inferred estimate of the median provides a good estimation of the true median, which is the key quantity in our analysis.

In summary, the kinetic energy gained through stellar encounters during one disk crossing can be defined by
\ben
\label{eq:dE_eff}
\Delta E &\simeq& {\rm Med}(\Delta E) = \kappa\, \Nc \frac{\overline{\delta v^2}}{2}\\
&\simeq & \kappa\,  \frac{\Nc}{2} \int \dd \mstar  \, p_{\mstar}(\mstar) \int_{b_0(Q=0.2)}^{b_{\rm max}} \dd b \, p_b(b) (\delta \vv)^2 \,,\nn
\een
with $\kappa \sim 0.7$. Here $\overline{\delta v^2}$ has been set to the integral of \refeq{delta_v2_average} truncated from below by imposing the cutoff $b_{\rm min} \to b_0(Q=0.2)$. Then, the total impact of stellar encounters during several disk crossings on one subhalo can be evaluated by replacing $\Delta E_{\rm d}$ by ${\rm Med}(\Delta E)$ in \refeq{recursion}.

\subsection{Validity of the impulse approximation}
\label{ssec:Impulse_approx}
Before moving on to our final results, let us discuss the validity of the impulse approximation, upon which all of our calculations have rested so far. That approximation is valid as long as the typical orbital timescale of DM particles within the subhalo is shorter than the duration of the encounter with a star, $t_{\rm col}\sim b/v_{\rm r}$. For a DM particle at a given position $r$ inside a subhalo, the orbital frequency is given by $\omega(r)=\sigma_{\rm sub}(r)/r$. The encounter is therefore impulsive if $t_{\rm col}\,\omega(r)\,\ll 1$ everywhere in the subhalo. For an NFW cuspy halo profile, the orbital frequency diverges at $r=0$ so the impulse approximation necessarily breaks down at some finite $r>0$, but this radius might still be very small compared to the scale radius $r_{\rm s}$ of the structure. Using the maximal impact parameter defined in \refeq{bmax}, and the mean relative speed defined in \refeq{mean_relative_speed}, we find that $t_{\rm col}\,\omega(10^{-3}\,r_{\rm s})<1$, regardless of the mass of the subhalo (we fix the concentration using the mass-concentration relation from \citeref{Sanchez-CondeEtAl2014}). At 8 kpc in a MW-like galaxy, for a subhalo with mass $10^{-10}\,\rm M_\odot$, we find that $t_{\rm col}\,\omega(10^{-3}\,r_{\rm s})\simeq 1$ for $v_{\rm r}\simeq 20\,\rm km/s$. For a single encounter, using the distribution of \refeq{relative_speed_distribution}, we find that the probability for the relative speed to be less than $20\,\rm km/s$ is $0.02\%$. We conclude that the impulse approximation is valid for the overwhelming majority of encounters down to radii as small as $10^{-3}\,r_{\rm s}$.

The situation is quite different when considering the complementary effect of disk shocking \ie~the gravitational shocking induced by the smooth potential of the disk rather than that of each individual star. In that case, the encounter duration is the disk-crossing time $t_{\rm cross}\sim z_{\rm d}/v_{\rm z}$ where $z_{\rm d}$ is the typical scale-height of the disk and $v_{\rm z}$ the subhalo crossing speed along the perpendicular direction. Because $z_{\rm d}\gg b_{\rm max}$, the impulse approximation breaks down more often in this case and adiabatic invariance must be accounted for to get accurate results \cite{Weinberg1994,Weinberg1994a}. This was discussed earlier around \refeq{Delta_E_d}.

\subsection{Results and comparisons with previous work}
\label{ssec:comp}

\begin{figure}[!t]
\centering
\includegraphics[width = 0.49\textwidth]{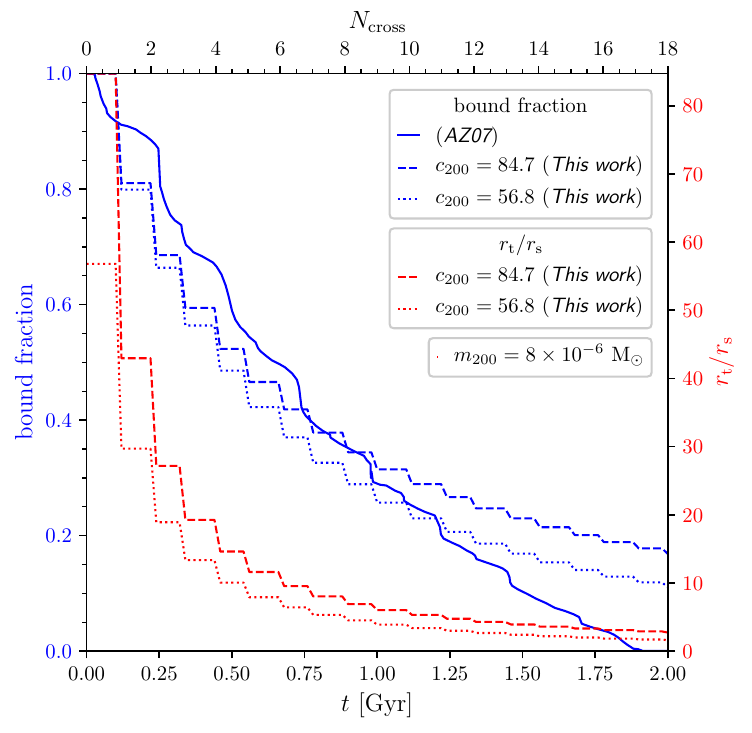}
\caption{\small Evolution of the mass fraction (blue) and tidal radius (red) for a subhalo of initial virial mass $8\times 10^{-6}$ $M_\odot$---virial mass considered in AZ07~\cite{AngusEtAl2007b}, and extrapolated to today. The initial subhalo radius is fixed to $r_{200}$. The concentration $c_{\rm 200} = 84.7$ corresponds to that of AZ07 extrapolated to today assuming fixed scale radius and fixed scale density. The concentration $c_{\rm 200} = 56.8$ is the median concentration from \citeref{Sanchez-CondeEtAl2014}.  }  
\label{fig:BoundMassFraction_vs_time}
\end{figure}

\begin{figure}[!t]
\centering
\includegraphics[width = 0.49\textwidth]{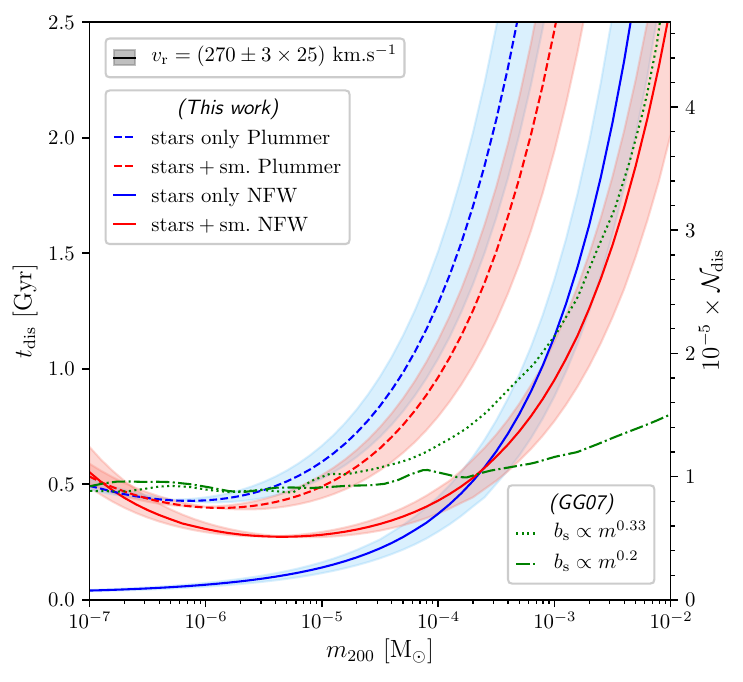}
\caption{\small Mass dependence of the disruption time associated with stellar encounters for a subhalo trapped at 8~kpc from the Galactic center in the stellar disk. Two cases are considered: (i) subhalos assumed with virial initial conditions (blue curves); (ii) subhalos with same virial parameters, but already pruned by tides induced by the smooth galactic gravitational potential (initial radii set to corresponding Jacobi radii). Comparison is made with GG07~\cite{GreenEtAl2007} reported as green curves, where the disruption time is defined from the fraction of total energy kick over the binding energy, see \refeq{DE_over_dU_behav}, and consider two asymptotic behaviors in $b$, with a transition at $b = b_{\rm s}$ (uncertainties on $b_{\rm s}$ are parameterized by two different mass dependencies).}  
\label{fig:DisruptionTime}
\end{figure}

\begin{figure}[!t]
\centering
\includegraphics[width = 0.49\textwidth]{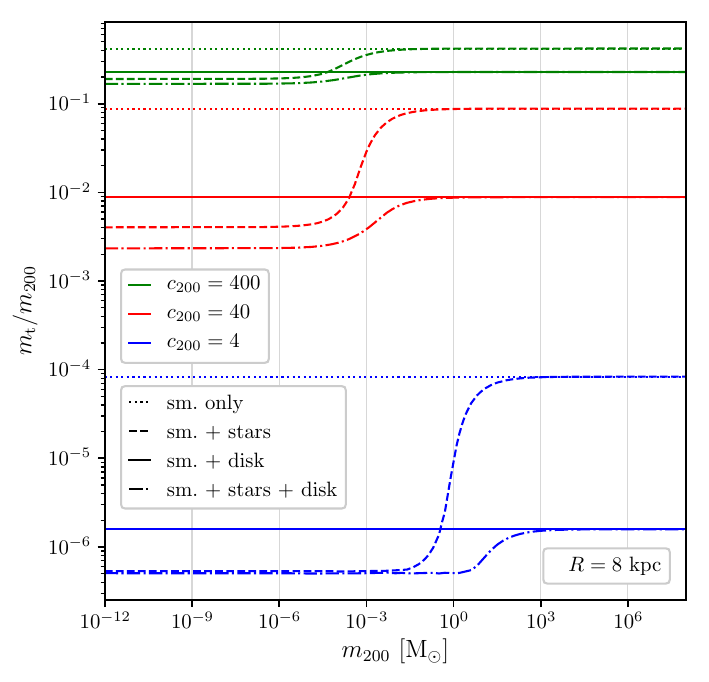}
\caption{\small Ratio of physical mass $m_{\rm t}$ to cosmological virial mass $m_{200}$ as a function of $m_{200}$ for several configurations of tidal effects, and three different concentrations. Are considered: smooth stripping only without baryons (dotted), stellar encounters (resp. disk shocking) on top of smooth stripping (dashed, resp. solid), all tidal effects (dash-dotted).} 
\label{fig:Mt_vs_m200}
\end{figure}

In \refig{BoundMassFraction_vs_time}, we show how the bound mass fraction and tidal radius $r_{\rm t}$ of a subhalo reduce in time because of stellar encounters. For comparison with complementary results from the literature, we consider a subhalo similar to that studied in the simulation of \citeref{AngusEtAl2007b} (hereafter AZ07)---their time-dependent bound mass fraction is reported as the solid blue curve. This subhalo is defined at redshift $z\sim 26$ with concentration $c_{\rm 200} = 2$ and virial mass $m_{\rm 200} = 10^{-6}$ M$_\odot$. To connect to our formalism, we rescale its size to $z\sim 0$, assuming a constant scale radius and scale density (which should be a valid approximation \cite{DiemerEtAl2019}). We find a concentration $c_{200}=84.7$, a mass $m_{200} = 8\times 10^{-6}$ M$_\odot$, and a virial radius $r_{200} = 0.42$ pc. We also show the case of a subhalo with the same virial mass but a concentration set to the median value picked in \citeref{Sanchez-CondeEtAl2014}. For simplicity, we consider that subhalos enter the galactic disk with an inclination $\cos\theta = 0.5$, and have a relative velocity with the stars $\overline{v_{\rm r}}(8 \, {\rm kpc}) = 334$ km$\cdot$s$^{-1}$. Our results are in good agreement with those of AZ07, derived from N-body simulations. The main difference is that, in our framework, stripping becomes less and less efficient with time. This discrepancy is mostly due to the fact that we assume a sharp tidal truncation radius while keeping the inner density profile unchanged at each crossing, which is not fully realistic \cite{StenDelos2019,GoerdtEtAl2007} but is hard to account for in an analytical subhalo population model---see further discussion on this in \citeapp{app:pdf_ccdf_DeltaEtilde}. In addition, note that while we derive an analytical estimate for the number of stellar encounters per crossing and the number of crossings, the authors of AZ07 use a more detailed galactic model.

In \refig{DisruptionTime}, we show the time $t_{\rm dis}$ it takes to completely destroy a subhalo with a mass $m_{200}$ (and median concentration) trapped in a MW-like galactic disk at 8~kpc from the halo center. We also indicate, on the right vertical axis, the corresponding number of encountered stars, $\Nc_{\rm dis}$, before disruption.  For comparisons with \citeref{GreenEtAl2007} (hereafter GG07), we consider both NFW and Plummer density profiles for the subhalo, and we assume stars with a relative velocity in the range $v_{\rm r} \sim (270\pm 3\times 25)$ km.s$^{-1}$. We focus on two cases: (i) the subhalo is taken with its virial parameters in the initial condition (radial extent of $r_{200}$ at $z \sim 0$); (ii) the subhalo is taken with the same virial parameters, but with a tidal radius $r_{\rm t}<r_{200}$ set to the Jacobi radius determined from the galactic potential at its position -- see \citeeq{eq:rt_smooth}. Comparing case (i) to GG07, we observe that the general behavior and the magnitude of $t_{\rm dis}$ are similar. Even if it does not impact the conclusion, let us point out, nevertheless, that the comparison can somewhat be biased. Indeed, the authors of GG07 considered subhalos at $z\sim 26$, similarly to AZ07. A mismatch between the definition of mass and virial radius for the same substructure could then arise. Furthermore, in our case, we assume that the gravitational potential of the subhalo does not change over the entire time it stays within the disk. This may artificially lower the value of $t_{\rm dis}$. Nonetheless, we show that it takes less time to destroy subhalos with an NFW profile than those with a Plummer profile (assuming the same mass). 

In that disruption analysis, we have considered resilient subhalos by choosing a disruption efficiency of $\epsilon_{\rm t} = 10^{-2}$. However, this choice appears not to have a strong impact. Indeed, when the number of encountered stars becomes large, there is a rapid transition between two distinct behaviors of the energy kick as a function of radius -- see \refeq{dE_eff}. Owing to the inverse dependence of $b_0(Q)$ with $\Nc$, see \refeq{b0}, it goes from $\Delta E \propto (r/r_{\rm s})^2$, as shown in \refig{MC_results_vs_r}, to $\Delta E \propto {\rm cst.}$, larger than the gravitational potential. Therefore subhalos experience a rapid transition with $\Nc$, between being only truncated (with a new tidal radius $r_{\rm t}$ close to the initial boundary), and being fully disrupted.

\section{Impact of stars on the subhalo population}
\label{sec:StellarPop_SubhaloModel}

\begin{figure*}[!t]
\centering
\includegraphics[width = 0.8\textwidth]{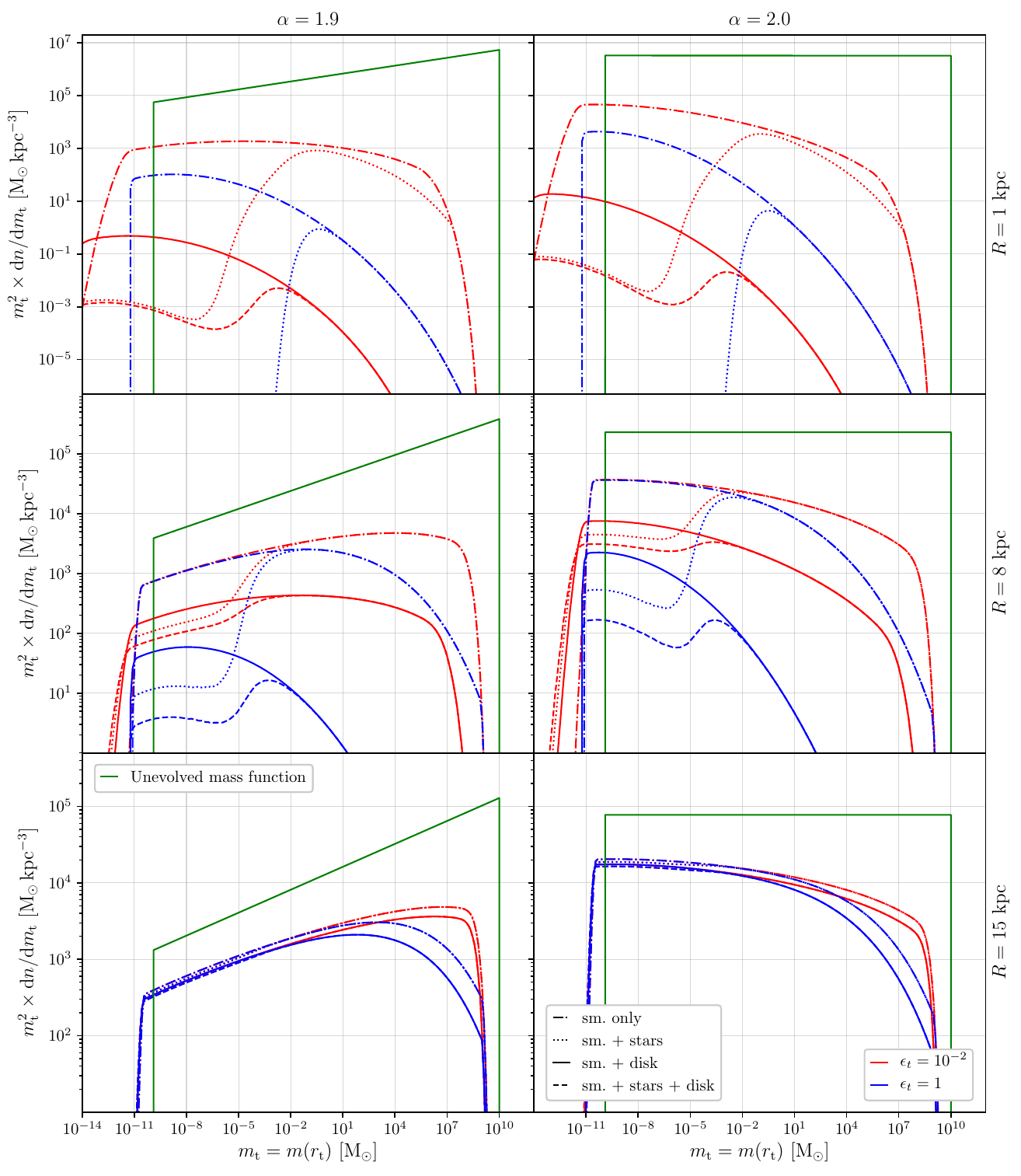}
\caption{\small Mass functions taking into account four different stripping configurations, at different distances from the Galactic center, $R$= 1, 8, 15~kpc, and for two different values of the mass index $\alpha =2.0$ on the left and $\alpha = 1.9$ on the right. The minimal cosmological mass is set to $m_{200}^{\rm min} = 10^{-10}$ M$_\odot$. The red (resp. blue) curves correspond to the resilient (resp. fragile) subhalos configuration. The green curve is the unevolved mass function for which $r_{\rm t} = r_{200}$. Notice that at $R=1$~kpc, there is no evolved fragile-subhalo mass function available when disk-shocking effects are plugged in, as they have destroyed all objects.} 
\label{fig:MassFunction}
\end{figure*}

\begin{figure*}[!t]
\centering
\includegraphics[width = 0.95\textwidth]{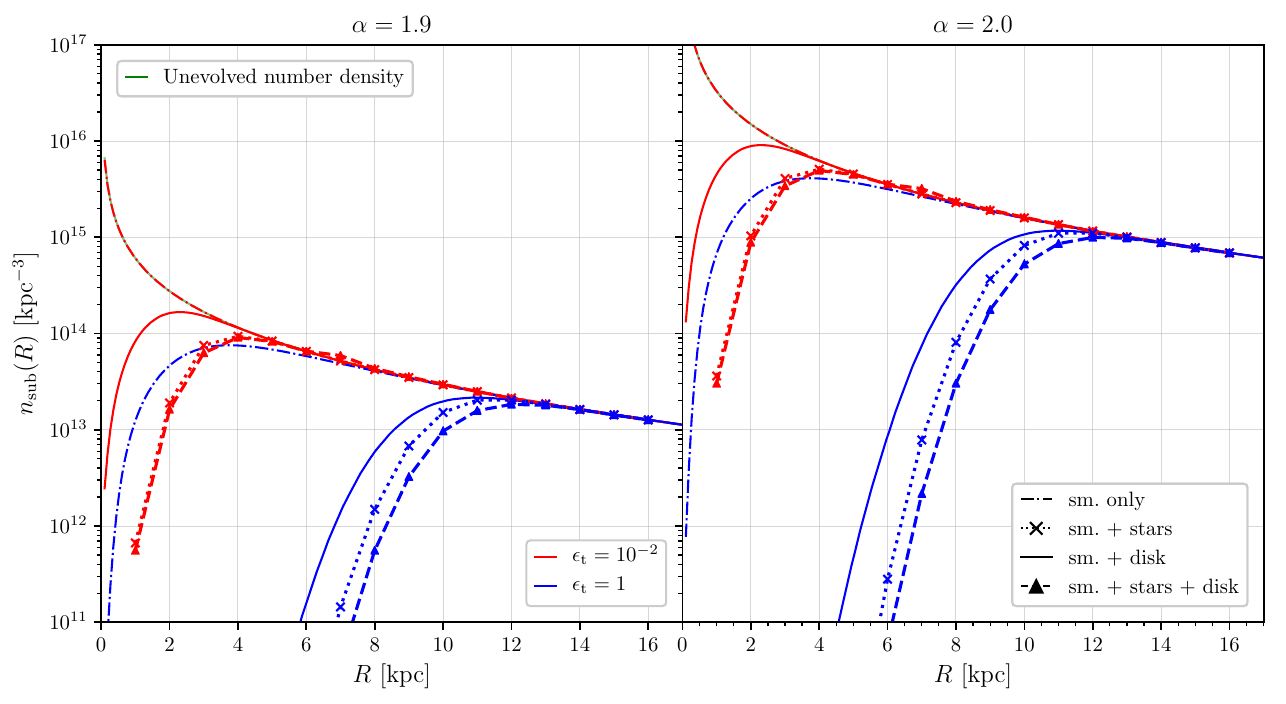}
\caption{\small Subhalo number density assuming four different stripping configurations, at different distances from the Galactic center, $R$= 1, 8, 15~kpc and for two different values of the mass index $\alpha =2.0$ (left panels) and $\alpha = 1.9$ (right panels). The minimal cosmological mass is set to $m_{200}^{\rm min} = 10^{-10}$ M$_\odot$. The red (resp. blue) curves correspond to the resilient (resp. fragile) subhalo configurations. The green curve is the cosmological number density in the hard-sphere approximation (without tidal effects). Notice that at $R=1$ kpc, disk-shocking effects destroy the entire population if subhalos are fragile.} 
\label{fig:NumberDensityStars}
\end{figure*}

In this section, we incorporate the effect of stellar encounters into the SL17 model in addition to those of smooth stripping and disk shocking. We start by describing how we combine the effects of individual encounters and disk shocking. Then, in a second step, we show our results for the impact on the subhalo mass function and on the total number density.

\subsection{Combination of the different stripping effects}
In the impulsive approximation limit, the total energy kick resulting from both stellar encounters and disk shocking can be merely written as the combination $\Delta \vv +  \Delta \vv_{\rm d}$, for which a PDF could be formally derived. However, because of adiabatic corrections in disk-shocking calculations (departure from the impulsive condition), such a derivation breaks down. Nevertheless, we can always write the total energy gain as
\ben
\Etot  &=& \frac{1}{2}(\Delta \vv + \Delta \vv_{\rm d})^2 + \vv \cdot (\Delta \vv + \Delta \vv_{\rm d}) \\
&=& \Delta E + \Delta E_{\rm d} +  \Delta \vv \cdot \Delta \vv_{\rm d} \, .\nn
\een
To cope with our ignorance of the true distribution of $\Delta \vv_{\rm d}$ once adiabatic corrections are included, we make the assumption that $ \Delta \vv \cdot \Delta \vv_{\rm d}  \sim 0$. This approximation is well-justified in the case of a subhalo with a normal incidence on the disk. Indeed, in the 2D random-walk scenario, $\Delta \vv$ is parallel to the disk and $\Delta \vv_{\rm d}$ is perpendicular to it. Therefore, we can approximate the total energy kick as
\ben
\Etot \simeq  \frac{1}{2} \[ 0.7 \Nc \overline{\delta v^2}  + |\Delta \vv_{\rm d}|^2A_1(\eta_{\rm d})\] \, .
\een
More details on the total distribution of $\Etot$ are given in \citeapp{app:pdf_ccdf_DeltaEtilde}, which support this definition. To speed up the calculations, $\overline{\delta v^2}$ is evaluated assuming a {\em typical} subhalo entering the galactic disk with an average inclination $\cos\theta = 0.5$, and an average relative velocity with the stars $\overline{v_{\rm r}}(R)$ given by \refeq{mean_relative_speed}. In the following, we recompute the tidal radii of subhalos by replacing the value of the kinetic energy kick in \refeq{recursion}, which only includes the smooth stripping and disk-shocking effects, by the above definition. We then show to what extent tidal stripping is impacted by stellar encounters.

\subsection{Results}
\label{sec:StellarPop_Results}
We can now present our final and complete results for the tidal stripping calculation (including disruption) at the level of a subhalo population. We consider four different configurations for the evaluation of the tidal effects. (i) {\em Smooth only}: the tidal radius is entirely defined by the Jacobi radius \refeq{rt_smooth}). (ii) {\em Smooth+stars}: on top of the smooth stripping, only individual encounters with stars are included. (iii) {\em Smooth+disk:} same as (ii), but only the disk-shocking effect is included. (iv) {\em Smooth+stars+disk:} all effects are included.  In \refig{Mt_vs_m200}, we show the final tidal subhalo mass in terms of the original cosmological mass for the different stripping configurations and three different concentrations, at a distance $R=8$ kpc from the Galactic center. The dominant effect of baryonic tidal stripping on small subhalos with initial mass $m_{200} \lesssim 1$ M$_\odot$ originates in individual stellar encounters. In contrast, for more massive subhalos, baryonic stripping is mostly due to the disk shocking.

The total mass function for the different stripping cases are displayed in \refig{MassFunction} for a minimal cosmological mass $m_{200}^{\rm min} = 10^{-10}$ M$_\odot$, and assuming two different initial spectral mass indices $\alpha$ (\ie~the slope of the initial cosmological mass function, see \citeapp{app:SL17}) in the SL17 model. While baryonic effects are mild and even negligible in the outer regions of the disk, for example at a distance $R \geqslant 15$~kpc, they have more and more impact toward the Galactic center. At $R=1$~kpc, the mass functions are strongly suppressed and shifted toward small masses, especially because of stellar encounters. Indeed, in the resilient subhalo scenario (disruption efficiency $\epsilon_{\rm t} = 10^{-2}$), they reduce the mass function by $\sim 6$ orders of magnitude for $10^{-10} < m_{\rm t} \lesssim 10^{-6}$ M$_\odot$ (with respect to the smooth-only case), and populate a mass range much below the minimal cosmological mass. Disk-shocking effects only produce an equivalent reduction of $\sim 4$ orders of magnitude. In the fragile case ($\epsilon_{\rm t} = 1$), disk-shocking effects disrupt all subhalos and so do stellar encounters at low masses. At $R=8$~kpc, we notice a similar effect with an almost $\sim 2$ (resp. 4) order-of-magnitude suppression due to stellar encounters and a $\sim 1$ (resp. 2) order-of-magnitude suppression due to disk shocking at low masses for resilient (resp. fragile) subhalos. The causes for the strength of stellar encounters in the center are twofold: close to the Galactic center, subhalos cross the disk more often and the stellar density is higher, reducing the interstellar distances and impact parameters, therefore enhancing the kinetic energy kicks.

In \refig{NumberDensityStars}, we show the subhalo number density as a function of the distance from the Galactic center. Farther than 12~kpc away from the Galactic center, baryonic tidal effects are negligible. Below $\sim$12 ~kpc, stellar encounters strongly impact the lighter subhalos, which are also the most numerous. Around 8~kpc, the effect is already significant. Although the number of subhalos in a resilient population is not impacted, a fragile subhalo population experiences a reduction of its number density by about one order of magnitude, compared with the impact of disk shocking and smooth stripping only (dashed and solid curves). The difference grows toward the Galactic center, and at $R =1$~kpc, the fragile subhalo population is almost  entirely destroyed by stellar encounters. At this distance, even the number density of the resilient subhalo population is further reduced by two orders of magnitude. These conclusions are valid for the two mass indices considered.

\section{Discussion and conclusion}
\label{sec:concl}
In this work, we have presented a theoretical, analytical analysis of the tidal stripping experienced by DM particles inside subhalos, and induced by baryonic components of galaxies. We have discussed two effects: disk shocking, for which baryons act collectively (and for which we summarized and slightly updated the results of previous developments \cite{StrefEtAl2017}), and individual stellar encounters, which we investigated in more detail. For the latter, in particular, we went beyond the reference work of Gerhard \& Fall \cite{GerhardEtAl1983} by studying penetrative encounters. We have derived a new analytical solution for the velocity kick received by particles in every radial shell of the subhalo in the impulsive approximation. We have then studied the impact of successive stellar encounters for a subhalo crossing the stellar disk from a thorough statistical analysis. Our results can be easily implemented in analytical or numerical models of subhalo populations \cite{BerezinskyEtAl2003,Benson2012,BerezinskyEtAl2014,HiroshimaEtAl2018,AndoEtAl2019}. They can be seen as complementary to studies carried out with numerical simulations \cite{GoerdtEtAl2007,StenDelos2019}.

A specific attention was paid to the cumulative energy kick received by particles in every radial shell of the subhalo, and we have performed Monte-Carlo simulations to cross-check our analytical estimates. These simulations show that a careful treatment of the impact parameter distribution is necessary not to overestimate the final kinetic energy, because very close but very unlikely encounters tend to falsely dominate in the naive calculation, and therefore to nonphysically bias averaged quantities. MC simulations also evidence that the energy-kick distribution is broad. This means that the median energy kick might not be a very reliable estimate of the energy received by a single subhalo. However, it can still be used to gauge the effect on an overall galactic subhalo population.

We have computed the subhalo mass function for a MW-like galaxy using the SL17 analytical subhalo model, and shown that stellar encounters have sizable effects in the inner 10~kpc on subhalos with masses $\lesssim 1$~M$_\odot$. This mass selection differs from other tidal interactions (smooth tides and disk shocking) that prune subhalos based on their concentration.

We have also implemented a simple tidal efficiency criterion to decide whether subhalos can survive tidal stripping, based on how deep toward their centers DM can be stripped away. If subhalos are fragile, as found in early cosmological simulation results \cite{HayashiEtAl2003}, their number density is strongly depleted by stars (the low-mass tail of the mass function below $\sim 1$~M$_\odot$ is efficiently depleted). On the other hand, if subhalos are resilient, as suggested by theoretical arguments and more recent dedicated simulation studies \cite{vandenBosch2017,vandenBoschEtAl2018,vandenBoschEtAl2018a,ErraniEtAl2020,ErraniEtAl2021}, the mass function is shifted to lower masses.

A caveat of our calculation is the assumption that inner subhalo profiles are preserved and that stellar encounters simply induce a sharp truncation at the induced tidal radius. In fact, we expect relaxation to modify the internal structure after each encounter \cite{HayashiEtAl2003, PenarrubiaEtAl2008, StenDelos2019}. While such inner DM redistribution could in principle be considered, its implementation would be challenging at the level of a full subhalo population in our analytical model without shattering its virtues in terms of calculation time -- see further discussion in \citeapp{app:ImpactOnProfile}. Nevertheless, that simple assumption remains in reasonable qualitative agreement with complementary numerical studies on stellar encounters (\eg~\citeref{AngusEtAl2007b}).

We also describe stellar encounters and disk shocking as independent from each other, while they are in fact two sides of the same coin. Disk shocking accounts for the average disk potential while individual star shocking accounts for the disk granularity. A more detailed picture can be obtained through the stochastic formalism \cite{Chandrasekhar1941,Chandrasekhar1942,Kandrup1980,Penarrubia2018,Penarrubia2019}. However, this approach is again more involved, and could hardly be incorporated in our semi-analytical subhalo population model.

In summary, the pruning of subhalos by stars can have a substantial impact on the prospects for (local) DM searches, especially if the latter are conducted in a stellar-rich environment. For instance, the probability that a subhalo passes through the Earth and enhances the local density by a non-negligible amount changes, which may be important for DM direct detection experiments \cite{IbarraEtAl2019}, or for other subhalo searches \cite{AdamsEtAl2004}. Moreover, if DM self-annihilates, the presence of subhalos boosts the local annihilation rate \cite{BergstroemEtAl1999a,PieriEtAl2011,LavalleEtAl2012,BartelsEtAl2015,AndoEtAl2019}, and this efficient pruning (if not disruption) expected in regions over which the stellar disk extends, should also modify the amplitude of this boost. Final, DM stripped away from subhalos should form dark streams \cite{ZhaoEtAl2007}. A quick estimate shows that, for $m_{\rm min} = 10^{-10}$ M$_\odot$, $10^{3}$ to $10^{5}$ subhalos should have crossed the Solar System and may have formed streams in the last 10 Gyr. It is therefore quite conceivable that the solar system be surrounded by a large amount of them. If detectable, they would give an interesting probe of the DM fine-grained structuring \cite{LisantiEtAl2012,KuhlenEtAl2012}. Eventually, one could also consider the heating of stars as a potential signature of the presence of subhalos \cite{Carlberg2012,FeldmannEtAl2015,Petac2019}.

\begin{acknowledgments}
This work has been supported by funding from the ANR project ANR-18-CE31-0006 ({\em GaDaMa}), from the national CNRS-INSU programs PNHE and PNCG, and from European Union's Horizon 2020 research and innovation program under the Marie Sk\l{}odowska-Curie grant agreement N$^{\rm o}$ 860881-HIDDeN --- in addition to recurrent funding by CNRS-IN2P3 and the University of Montpellier. GF acknowledges support of the ARC program of the Federation Wallonie-Bruxelles and of the Excellence of Science (EoS) project No. 30820817 - be.h “The H boson gateway to physics beyond the Standard Model” 
\end{acknowledgments}

\bibliographystyle{apsrev4-2}
\bibliography{Stellar_encounters}

\clearpage

\appendix

\begin{widetext}

\section{The SL17 subhalo population model: a statistical semi-analytical model}
\label{app:SL17}

The SL17 model \cite{StrefEtAl2017} was motivated by the need to build a global galactic mass model including both a smooth DM component and a subhalo population aside from baryonic components, easy to make consistent with potential observational constraints, and in which tidal effects would be calculated from the very components of the model itself. The main constraint that was imposed from the beginning was that the sum of the smooth DM density profile and of the smoothed overall density profile of the subhalo component should give the global DM halo profile, \ie\ the one that can be constrained from observational data (see \eg\ \citeref{McMillan2017}).

In the SL17 model the mass density of each subhalo is described by its inner profile. In this study we use a standard NFW profile if not said otherwise. We also consider a Plummer \cite{Plummer1911} profile for some applications. The mass density at a distance $r$ from the center of the subhalo can be parameterized, in both cases, under the form
\ben
\rho\(x\equiv r/r_s\) = \rho_{\rm s} \,x^{-\gamma_\rho}\[1+ x^{\alpha_\rho}\]^{\frac{\gamma_\rho- \beta_\rho}{\alpha_\rho}} \, , 
\een
with $\rs$ and $\rhos$  the scale radius and the scale density respectively, and $x$ the dimensionless radius. In the NFW case, $(\alpha_\rho, \beta_\rho, \gamma_\rho)=(1, 3, 1)$ while in the Plummer case $(\alpha_\rho, \beta_\rho, \gamma_\rho)= (2,5,0)$. Henceforth, a subhalo is characterized by three quantities, $\rs$, $\rhos$ as well as its distance $R$ from the Galaxy center  -- circular orbits are assumed. Conveniently, it is also possible to describe the profile from cosmological parameters: the virial mass and concentration. The virial mass, denoted $m_\Delta$, corresponds to the mass contained inside a radius $r_\Delta$ over which the subhalo has an average density equal to $\Delta$ times the critical density $\rhocrit= 3 H_0^2/(8\pi G)$, with $H_0=67.4$ km/s/Mpc, the Hubble parameter \cite{PlanckCollab2020}. This yields $m_\Delta = (4/3)\pi r_\Delta^2 \Delta \rhocrit$. The concentration is defined by $c_\Delta = r_\Delta/r_{\rm s}$, and there is a one to one relationship between the couples $(m_\Delta, c_\Delta)$ and $(\rho_{\rm s}, r_{\rm s})$. In practice the value  $\Delta = 200$ is used \cite{BryanEtAl1998} as it is a good approximation for the critical over-density of subhalos when they virialize, in the matter-dominated Universe. \\

Owing to several dynamical effects, subhalos are tidally pruned. Their physical tidal extension is not defined by the cosmological size $r_{200}$ they would have in a flat background, but by their tidal radius $r_{\rm t}$. According to results of cosmological simulations \cite{TormenEtAl1998,HayashiEtAl2003,DiemandEtAl2004,DiemandEtAl2008b,SpringelEtAl2008,vandenBosch2017} we expect subhalos that are stripped too much (\ie\ that have too small a tidal radius) to be destroyed. In the model, this is implemented by the criterion
\ben
\begin{cases} 
x_{\rm t}= r_{\rm t}/r_{\rm s} \ge \epsilon_{\rm t} \quad \Rightarrow \quad \text{the subhalo survives}\\
x_{\rm t}= r_{\rm t}/r_{\rm s} < \epsilon_{\rm t} \quad \Rightarrow \quad \text{the subhalo is disrupted}
\end{cases}
\een
that relies on the value of $\epsilon_{\rm t}$,  treated as a fixed constant input. The lower this coefficient is, the more resilient subhalos are to tidal stripping. Two values are considered in the following: $\epsilon_{\rm t}=1$ in agreement with cosmological simulation where subhalos are rather fragile, and $\epsilon_{\rm t}=10^{-2}$ following the semi-analytical studies of \cite{vandenBoschEtAl2018a,vandenBoschEtAl2018,ErraniEtAl2020,ErraniEtAl2021} where cuspy subhalos are shown to be instead very resilient to tides.\\

The SL17 model does not only describe individual sub-halos but their entire population using a joint probability distribution function (PDF) on the virial mass, concentration and position of all subhalos. Assuming that all clumps are independent from each other this global PDF can be factorized into $N_{\rm tot}$  one-point PDFs, with $N_{\rm tot}$ being the total number of surviving subhalos. The value of $N_{\rm tot}$ is normalized consistently against DM only, numerical simulations (more precisely the Via Lactea DM only results \cite{DiemandEtAl2008b}). The one point PDF is given as
\beq
\begin{split}
p_{\rm t}(m_{200}, c_{200}, R)  \equiv \frac{1}{K_{\rm t}}   p_{\Rv}(R)\,  p_m(m_{200})  p_c(c_{200} | m_{200})   \Theta \[\frac{r_{\rm t}}{r_{\rm s}}(m_{200}, c_{200}, R) - \epsilon_{\rm t}\]  
\end{split}
\eeq
where $K_{\rm t}$ is a normalization parameter to have a probability of one if integrated on the entire parameter space. The PDF for the position $p_{\Rv}$ is obtained considering that the subhalo spatial distribution follows the global profile of the total DM halo. The PDF for the mass is obtained trough cosmological mass function. Cosmological simulation exhibit power-law mass functions $\propto m^{-\alpha}$ with a mass index $\alpha \lesssim 2$ \cite{DiemandEtAl2008b,ZhuEtAl2017, SpringelEtAl2008,DiemandEtAl2007,DiemandEtAl2006}. This is theoretically backed-up by the Press-Schechter formalism and its extension \cite{BondEtAl1991a,PressEtAl1974,LaceyEtAl1993,LaceyEtAl1994,ShethEtAl2001}, even if the small mass range is still weakly constrained today. Therefore we set $p_m(m_{200}) \propto m_{200}^{-\alpha}$ with $\alpha \in [1.9, 2.0]$. Besides, the virial mass must be bounded from below by $m_{200}^{\rm min}$, here set as a free parameter of the model. Within a thermal DM particle model the minimal mass is fixed by kinetic decoupling in the Early Universe \cite{BringmannEtAl2007a,Bringmann2009,GreenEtAl2005,Bertschinger2006a,HofmannEtAl2001} and can go down to $10^{-12}$ M$_\odot$. Eventually, the PDF for the concentration is a log-normal \cite{WechslerEtAl2002,MaccioEtAl2007,MaccioEtAl2008,Jing2000,BullockEtAl2001}, whose median, given in Ref.~\cite{Sanchez-CondeEtAl2014}, is fitted against numerical simulations.  The Heaviside function $\Theta$, which encodes the subhalo disruption, leads to the entanglement of the latter three PDFs through the dependency of $r_{\rm t}$ on $m_{200}$, $c_{200}$ and $R$. The full PDF $p_{\rm t}$ is therefore a complicated, non-separable function.\\

With this formalism, it is possible to describe more precisely the decomposition of the density of the total DM halo $\rho_{\rm tot}$, as the sum 
\ben
\rho_{\rm tot}(R) = \rho_{\rm sm}(R) + \int \dd m_{\rm t}  \, m_{\rm t} \der{n_{\rm sub}}{m_{\rm t}}(R)
\een
where  $\rho_{\rm sm}$ the density of DM in the smooth component of the halo and the integral corresponds to the contribution of subhalos. The function ${\rm d} n_{\rm sub}/{\rm d} m_{\rm t}$ is the local evolved (\ie{} after stripping) subhalo mass function. It is related to the subhalo PDF through
\beq
\begin{split}
\der{n_{\rm sub}}{m_t}(R)  =   N_{\rm tot} \int  \dd c_{200} \, \dd m_{200} \, p_{\rm t}(m_{200}, c_{200}, R)  \delta_{\rm D}\[m_{\rm t} - m\[r_{\rm t}(m_{200}, c_{200}, R)\]\] \, ,
 \label{eq:massFunction}
\end{split}
\eeq
where $m(r_t)$ is the mass of the subhalo within the tidal radius $r_{\rm t}$. Note that $m_{\rm t}$ is the tidal mass of the subhalo \ie{}\ its physical mass as opposed to its cosmological virial mass $m_{200}$. Tidal effects come into play through $p_{\rm t}$ (which encodes whether subhalos are destroyed or not) and directly through the tidal radius $r_{\rm t}$. The next section is dedicated to the evaluation of $r_{\rm t}$ at different radii, masses, and concentrations as implemented in SL17.

\section{Velocity dispersion}
\label{app:velocity_pdf}
To compute the velocity dispersion $\sigma$, we start from the Jeans equation for a spherical system,
\ben
\frac{1}{\rho}\frac{\partial(\rho\left<v_r^2\right>)}{\partial r}+2\,\frac{\beta}{r}\,\left<v_r^2\right> = -\frac{\dd\Phi}{\dd r}\,,
\quad {\rm where} \quad 
\Phi(r) \equiv -\int_{r}^{r_{\rm t}}\dd r'\,\frac{G_{\rm N}\,m(r')}{r'^2}
\label{eq:def_gravitational_potential}
\een
is the gravitational potential with $r_{\rm t}$ the tidal radius of the subhalo, and 
\ben
\beta(r) \equiv 1-\frac{\left<v_\theta^2\right>+\left<v_\phi^2\right>}{2\,\left<v_r^2\right>}
\een
is the anisotropy parameter. Here isotropy is assumed, therefore we have $\beta=0$ and $\sigma_{\rm sub}^2 = \left<v_r^2\right>$ thus
\ben
\sigma_{\rm sub}^2(r) = \frac{G_{\rm N}}{\rho}\int_{r}^{r_{\rm t}}\dd r'\,\frac{\rho(r')\,m(r')}{r'^2}\,.
\een
The same approach is used to compute the velocity distribution of subhalos in the dark halo of the Galaxy. One major difference is that baryons now contribute to the potential $\Phi$ and the mass of the system is 
\ben
m_{\rm tot}(R) =m_{\rm DM}(R) + \int_{|\rv'| < R } \dd^3 \rv' \,\rho_b(\rv')\,  .
\een
where $\rho_{\rm b}$ is the baryonic mass density, which is axisymmetric rather than spherical. The DM velocity variance is then
\ben
\sigma^2(R) = \frac{G_{\rm N}}{\rho}\int_{R}^{R_{\rm max}}\frac{\rho(R')\,m_{\rm tot}(R')}{R'^2}\,\dd R'\,,
\een
where the radial extension of the dark halo is fixed to $R_{\rm max}=500\,\rm kpc$.


\section{The stellar disks}

We use the mass model of McMillan \cite{McMillan2017}, in which two stellar disks parameterized by the axisymmetric density function
\ben
\rho_{\rm d}(R,z) = \frac{\Sigma_0}{2\,z_{\rm d}}\,\exp\left(-\frac{R}{R_{\rm d}}-\frac{|z|}{z_{\rm d}}\right)
\een
are fitted against a number of observational constraints. The best-fit parameters are shown in \citetab{tab:disk_parameters}
\begin{table}[!h]
\begin{tabular}{c|ccc}
 & $\Sigma_0$ & $R_{\rm d}$ & $z_{\rm d}$ \\
 & ${\rm [10^8\,M_\odot/kpc^2]}$ & ${\rm [kpc]}$ & ${\rm [kpc]}$ \\
 \hline
 thin & 8.96 & 2.5 & 0.3 \\
 thick & 1.83 & 3.02 & 0.9 \\
\end{tabular}
\caption{\small Stellar disk parameters.}
\label{tab:disk_parameters}
\end{table}

\section{Distribution of impact parameters}
\label{app:FullCompDistribImpactParam}

\begin{figure}[!t]
\centering
\includegraphics[width = 0.5\textwidth]{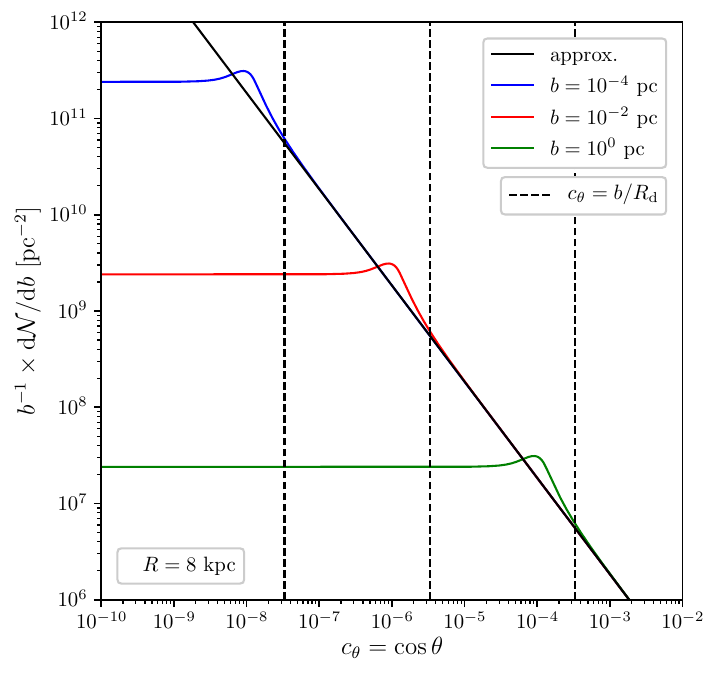}
\caption{\small Number density of stars encountered with an impact parameter $b$ divided by $b$ (to ease the comparison between the approximation and the full expression) vs. the cosine of the angle between the subhalo trajectory and the normal to the galactic plane $\theta$. The approximate calculation is given by the black solid curve and is independent of $b$, while the full expression gives the colored solid curves for different values of $b$. There is a change in behavior around $\cos \theta \sim b/R_{\rm d}$ where $R_{\rm d} \sim 3$ kpc is the approximate length scale of a MW-like stellar disk.}  
\label{fig:ImpactParam}
\end{figure}

We define the galactic frame with its origin at the galactic center $(G, \ev_x, \ev_y, \ev_z)$, the $z$-axis perpendicular to the galactic plane (in the following, the galactic plane refers to the middle of the stellar disk), and $\ev_x$ and $\ev_y$ arbitrary orthonormal vectors defining the galactic plane. We assume that during disk crossing, a subhalo keeps a rectilinear trajectory with a velocity $\vv_c$ that intercepts the galactic plane at one point, at position $\Rv$ at time $t=0$ in the galactic frame. Consider now a star tagged by the letter $i$ that has a position $\Rv_i$ at $t=0$ in the same frame. We assume that during the time interval it takes for the subhalo to cross the disk, that star keeps a linear trajectory with a velocity $\vv_i$ such that $\vv_i \cdot \Rv_i \sim 0$. For simplicity, we also introduce ${\bf S}_i = \Rv_i - \Rv$. Then, at an arbitrary time $t$, the distance between the star and the subhalo is given by
\ben
d_i(t) = | \Sv_i - \vv_{{\rm r},i} t |\,,
\een
where we introduce the relative velocity between the star and the subhalo $\vv_{{\rm r},i} = \vv_c - \vv_i$.  The impact parameter for this specific star is defined as $b_i \equiv {\rm min} \{d_i(t)\}$, which yields
\ben
b_i = \frac{|\Sv_i \times  \vv_{{\rm r},i}|}{| \vv_{{\rm r},i}|} \, .
\een
We want to determine the probability for this star to be at position $\Rv_i$ and have a mass $m_i$. We call the associated probability distribution $p_{(\mstar, \Rv_\star)}(m_i, \Rv_i) =  p_{\mstar}(m_i \, |\, \Rv_i) p_{\Rv_\star}(\Rv_i) $. In our model we make the usual approximation that  $p_{\mstar}(m_i \, |\, \Rv_i) =  p_{\mstar}(m_i)$, so that it does not depend on the position. Moreover, the probability distribution of positions is given by the mass density as
\ben
p_{\Rv_i}(\Rv_i) = \rho_\star(\Rv_i) \[\int \dd^3 \Rv \, \rho_\star(\Rv)\]^{-1} = \frac{\rho_\star(\Rv_i)}{\mstar^{\rm tot}}\,,
\een
where $\mstar^{\rm tot}$ is the total mass in stars, and $\rho_\star(\Rv_i)$ the associated mass density at position $\Rv_i$. All stars being independent, the joint PDF for their mass and position is
\ben
p_{(\mstar, \Rv_\star)}\(\{m_i\}_i, \{\Rv_i\}_i\)  =  \prod_{i=1}^{N_\star} \bigg[ p_{\mstar}(m_i)  p_{\Rv_\star}(\Rv_i) \bigg]\,,
\een
with $N_\star = m^{\rm tot}_\star/\overline{\mstar} $ the total number of stars. With all these ingredients, it is possible to evaluate the number density of stars that are encountered with impact parameter $b$ and mass $\mstar$, knowing the trajectory of the subhalo. It is given by
\ben
\der{^2\Nc}{b \dd \mstar}= \int \prod_{i=1}^{N_\star}  \dd m_i \dd^3 \Rv_i \,   p_{\mstar}(m_i)p_{\Rv_\star}(\Rv_i) \[\sum_{i=1}^{N_\star} \delta_{\rm D}(b_i - b) \delta_{\rm D}(m_i - \mstar) \]\, .
\een
Massaging this expression, it is straightforward to prove that the mass probability distribution can be factored out. This leaves us with the number density of stars with a given impact parameter that can be written under the compact form
\ben
\der{\Nc}{b }=  \frac{1}{ \overline{\mstar}} \int \dd^3 \Rv_\star\,  \rho_{\star}(\Rv_\star) \delta_{\rm D} (b_\star - b)\, .
\label{eq:dMdb}
\een
Form the expression of the impact parameter, it is convenient to make the change of variable $\Rv_\star \to \Sv_\star = \Rv_\star - \Rv$ in this integral, and to define the Dirac distribution on the squared value of the impact parameter. Eventually, to simplify the computation, we assume that all stars lay within an infinitely thin axisymmetric disk of surface density $\Sigma_\star$, so that the integration over the entire 3D space reduces to the integration over the Galactic plane. We can therefore write
\ben
\der{\Nc}{b }=  \frac{2b}{ \overline{\mstar}} \int \dd^2 \Sv_\star\,  \Sigma_{\star}\[R_\star(\Sv_\star)\] \delta_{\rm D} (b^2_\star - b^2)\, .
\een
Now, without loss of generality, we can parameterize the problem by choosing the convenient orientation of the basis $(\ev_x, \ev_y)$ such that $\Rv = (R, 0, 0)$. Moreover, we define the relative velocity direction as $\vv_{{\rm r},i}/ | \vv_{{\rm r},i}|= (\sin \theta \cos\varphi, \sin\theta \sin \varphi, \cos \theta)$, and  $\Sv_\star= (S \cos \phi, S \sin\phi, 0)$ such that $\dd^2 \Sv_\star = s \dd s \dd \phi$. Then, the expression of the impact parameter becomes
\ben
b_\star^2 = s^2 \[1 - \sin^2 \theta \cos^2(\phi - \varphi)\] \, .
\een
The integration over the Dirac delta distribution in \refeq{dMdb} after the change of variable can then be done analytically by solving the delta for the angle $\phi$. This gives four distinct solutions in $[\varphi - \pi, \varphi + \pi]$ when $Sc_\theta < b < S$. These solutions can be written under the form
\ben
\phi_j = \varphi + \eta_j \arccos\[\frac{\chi_j}{\sin \theta}\sqrt{1-\frac{b^2}{s^2}}\] \, ,
\een
where we have introduced $\chi_j = (+1, +1,  -1, -1)$ and $\eta_j = (+1, -1, +1, -1)$. The arccos function image being in the range $[0, \pi]$ only, the $\chi_j$ factor parameterizes the two solutions in the interval $[\varphi, \varphi + \pi]$ while $\eta_j$ gives the two symmetric solutions in the interval $[\varphi-\pi , \varphi]$. Henceforth, in order to perform the integration over the Dirac distribution, it is also necessary to provide the absolute value of the derivative of $b_\star^2$ with respect to $\phi$, evaluated at each solution point. Using the properties of these solutions, this takes a simple form
\ben
\left| \der{b_\star^2}{\phi} \right|_{\phi = \phi_j} & = & 2 s^2 \sin^2\theta |\cos(\phi_j - \varphi)||\sin(\phi_j - \varphi)|  =  2 \sqrt{s^2 - b^2}\sqrt{b^2 - s^2\cos^2\theta} \, .
\een
Then, we need to relate the value of $R_\star$ to $s$ at each solution point. We denote $R_j$ these four quantities and introduce a new variable $y = s^2/b^2$. Using the simple relation $\Rv_\star = \Sv_\star + \Rv$ -- that is, the definition of $\Sv_\star$ -- it is straightforward to show that $R_j^2 =  R^2 + b^2y + 2h_j(y, c_\theta, \varphi)$, with the shorter notation $c_\theta = \cos\theta$ and the functions $h_j$,
\ben
h_j(y, c_\theta, \varphi)  \equiv \frac{\chi_j R b}{\sqrt{1-c^2_\theta}}   \[\cos\varphi \sqrt{y-1} - \frac{\eta_j}{\chi_j}\sin\varphi \sqrt{1-y c^2_\theta}\]\,.
\label{eq:Rj}
\een
In the end, the number density of stars with impact parameter $b$ can be written
\ben
\der{\Nc}{b } =  \frac{b}{2}  \int_{1}^{\frac{1}{c_\theta^2}}  \frac{\dd y }{\sqrt{1 - yc^2_\theta}\sqrt{y- 1}}  \sum_{j=1}^{4} \frac{\Sigma_\star(R_j)}{\overline{\mstar}} \,,
\een
and the computation is complete. Eventually we also need to relate the angles $\varphi$ and $\theta$ to physical quantities (\ie~the angles made between the directions of the stars and the subhalo). In practice, this is difficult in the general case. We therefore consider that stars are motionless during the whole disk crossing so that $\vv_{{\rm r}, i} = \vv_c$.  \\

Formally, functions $h_j$ can be rewritten in a more convenient way as
\ben
h_j = \chi_jRb\sqrt{y}\cos\(\varphi + \frac{\eta_j}{\chi_j} \arctan\[\sqrt{\frac{1-yc_\theta^2}{y-1}}\]  \)\,.
\een
Under this form, it becomes easier to see that the $h_j$ and the associated radii $R_j$ are bounded,
\ben
|h_j| \le Rb\sqrt{y} \quad {\rm and} \quad |R-b\sqrt{y}| < R_j < R+b\sqrt{y}  \, .
\label{eq:Bounds_Rj}
\een
In particular, since $y < 1/c_\theta^2$, this previous inequality implies that $|R_j - R|< b/c_\theta$. If we consider that the variations of $\Sigma_\star$ as a function of $R$ are of typical length $R_{\rm d}$, then this implies that $\Sigma_\star(R_j) = \Sigma_\star(R)$ as long as $c_\theta \gg b /R_{\rm d}$. Therefore we can simply rewrite the density of encountered stars as
\ben
\der{\Nc}{b }  =  2b \frac{\Sigma_\star(R)}{\overline{\mstar}}   \int_{1}^{\frac{1}{c_\theta^2}}  \frac{\dd y }{\sqrt{1 - yc^2_\theta}\sqrt{y- 1}} =  \frac{\Sigma_\star(R)}{ \overline{\mstar}}   \frac{2\pi b\dd b}{c_\theta} \, .
\een

\section{Energy and velocity distributions in stellar encounters}
\label{app:E_and_v_pdfs_for_encounters}
In this appendix section, we detail the PDFs of several functions of interest, and prove several properties used in the main text. We start by analytically deriving the PDF of final velocities in a given shell. Then we prove the relation between the median and the average kinetic energy kick of \refeq{RelationAverageMedian}. In a third part, we illustrate this formal derivation with the example of initial velocities following a Maxwell-Boltzmann distribution. Eventually, we conclude by studying the impact of stellar encounters on the density profile using a simple criterion in order to justify the use of the median as the typical kinetic energy kick felt by all particles in a given shell during one crossing of the disk.

\subsection{Probability distribution for the final velocity}
We suppose that we know the PDF, $p_{\vv}(\vv\, |\, r)$, of the initial velocity $\vv$ of particles at position $r$ in a subhalo, and also the probability for particles in that shell to receive a velocity kick $\Delta \vv$. The probability distribution of final velocity $v_f$ at position $r$ can therefore be written under the form
\ben
p_{v_f} (v_f\, | \, r)  =  \int \dd^3 \vv \,  p_{\vv}(\vv\, | \, r) \int \dd^d \Delta \vv \,  p_{\Delta \vv}(\Delta \vv\, | \, r) \delta_{\rm D}\[v_f - |\vv + \Delta \vv| \] \, .
\een
We leave room here for the possibility of $\Delta \vv$ being a 3D or 2D random vector with the dimension parameter $d$. Let us assume an isotropic initial velocity distribution so that $p_{\vv}(\vv \, | \, r) = p_{\vv}(v \, | \, r)$ with $v = |\vv|$. Then, it is possible to integrate first over the angular distribution of $\vv$ in order to remove unnecessary degrees of freedom, and to get rid of the Dirac delta term. This yields
\ben
p_{v_f} (v_f\, | \, r) = \int \dd^3 \vv  \, p_{\vv}(v \, | \, r)  p_{v_f}(v_f\, | \, v, r)\,,
\label{eq:pvf}
\een
with the definition of the PDF of $v_f$ (given $v$ and $r$) being
\ben
p_{v_f}(v_f\, | \, v, r) = \frac{v_f}{v}  \int \dd^d \Delta \vv \,  \frac{p_{\Delta \vv}(\Delta \vv\, | \, r) }{\Delta v}\Theta\[\Delta v - |v+v_f|\]\Theta\[(v+v_f) - \Delta v\]\,,
\een
and where we have introduced the velocity kick norm $\Delta v \equiv |\Delta \vv|$. For an isotropic distribution of $\Delta \vv$ such that $p_{\Delta \vv}(\Delta \vv \, | \, r) = p_{\Delta \vv}(\Delta v \, | \, r)$, it further simplifies as
\ben
p_{v_f}(v_f\, | \, v, r) = \frac{v_f}{v} \frac{\pi^{d/2}}{\Gamma(d/2)}  \int_{|v-v_f|}^{|v+v_f|} \dd \Delta v \, (\Delta v)^{d-2 } p_{\Delta \vv}(\Delta v\, | \, r) \, .
\een
Let us now assume that $\Delta \vv$ follows a Gaussian distribution according to the central limit theorem. Then the distribution on $\Delta \vv$ takes a simple form,
\ben
p_{\Delta \vv}(\Delta v\, | \, r) =  \frac{\pi^{-d/2}}{ u^d}e^{-\frac{(\Delta v)^2}{u^2}} \quad {\rm with} \quad u = u(r) = \sqrt{\frac{2\Nc \overline{(\delta \vv)^2}}{d}}\,,
\een
which gives, with the change of variable $\Delta v \to u\sqrt{t}$ in the second line, and the introduction of the incomplete Gamma function in the third line,
\ben
p_{v_f}(v_f\, | \, v, r)&  = & \frac{v_f}{v} \frac{1}{\Gamma(d/2)}  \int_{|v-v_f|}^{|v+v_f|} \dd \Delta v \, \frac{(\Delta v)^{d-2 }}{u^d} e^{-\frac{(\Delta v)^2}{u^2}} \\
 & =&  \frac{v_f}{u v} \frac{1}{2\Gamma(d/2)}  \int_{\(\frac{v-v_f}{u}\)^2}^{\(\frac{v+v_f}{u}\)^2} \dd t \,t^{\frac{d-3}{2}} e^{-t} \\
 & = &  \frac{v_f}{u v} \frac{1}{2\Gamma(d/2)}  \[\Gamma\(\frac{d-1}{2}, \(\frac{v-v_f}{u}\)^2\) - \Gamma\(\frac{d-1}{2}, \(\frac{v+v_f}{u}\)^2\)\] \, .
 \label{eq:finalVelocity}
\een
We have just found a generic expression for the PDF of $v_f$ for any initial velocity distribution. Another interesting quantity is the associated CDF of $v_f$ that is defined as
\ben
F_{v_f}(< v_f\, | \, r) & \equiv&  \int_{0}^{v_f} p_{v_f}(v_f'\, | \, r)\,  \dd v_f' =  \int \dd^3 v  \, p_{\vv}(v \, | \, r)  \left\{ F_{v_f}(<v_f\, | \, v, r) \equiv\int_0^{v_f} p_{v_f}(v_f'\, | \, v, r) \dd v_f'  \right\} \, .
\een
The CDF given $v$ and $r$ can therefore be computed with the expression of the PDF derived above. We will see the utility of this expression later below, especially when discussing the modifications to subhalo density profile due to the encounter. However, let us first inspect the properties of the median energy kick in light of this derivation.

\subsection{Properties of the median energy kick}

\begin{figure}[!t]
\centering
\includegraphics[width = 0.5\textwidth]{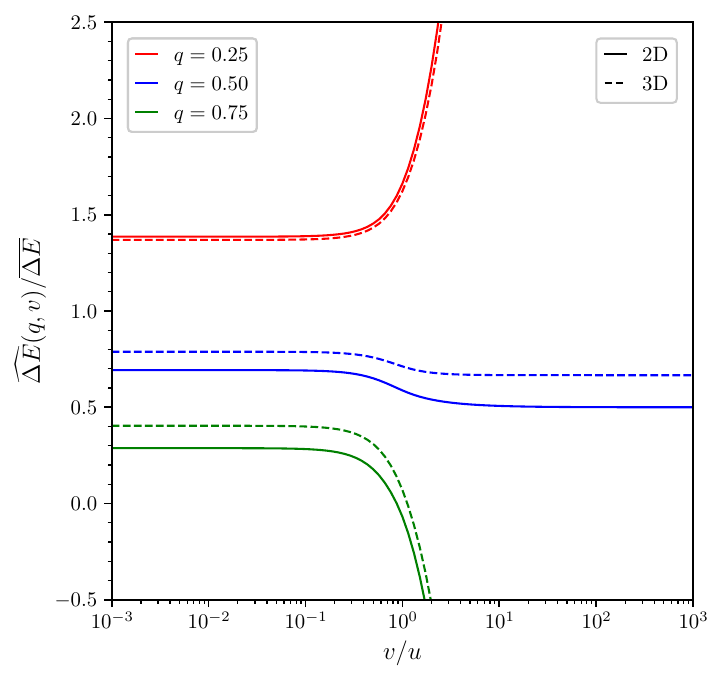}
\caption{\small The inverse CCDF of kinetic energy knowing the velocity at three values $q = 0.25$, $q = 0.5$ and $q=0.75$ with respect to the initial velocity $v$}  
\label{fig:InverseCCDF_DEv}
\end{figure}

In the main text we have introduced the median value of the energy kick received by particles in a subhalo during the encounter with stars, as it crosses the stellar disk. Moreover, we have based our computation of the tidal radius on the fact that the typical energy received in a shell is defined as the median value that can be approximated by the average value modulated by a coefficient between 0.5 and 0.7 (in the case of a 2D random walk in the velocity space). We prove this property here, and show that it is independent of the initial velocity distribution. Similarly to the CDF for $v_f$, it is possible to introduce a related CDF for the kinetic energy kick $\Delta E$ as follows
\ben
F_{\Delta E}(< \Delta E\, | \, r) =  \int \dd^3 v  \, p_{\vv}(v \, | \, r)  F_{v_f}(<\sqrt{v^2+2\Delta E}\, | \, v, r) \, .
\een
Henceforth, we can define a median value for $\Delta E$ given $v$ and $r$ as ${\rm Med}(\Delta E \, | \, v, r )$. In \refig{InverseCCDF_DEv}, we show that value of $\widehat{\Delta E}(q, v)$, which is given by the implicit equation 
\ben
F_{v_f}\(<\sqrt{v^2+2\widehat{\Delta E}(q, v)}\, | \, v, r\) = 1- q,
\een
for three different values of $q$. In particular, the median ${\rm Med}(\Delta E \, | \, v, r ) = \widehat{\Delta E}(0.5, v)$ is bounded by the asymptotes in $v = 0$ and $v \to \infty$. Series expansions in these two regime further show that the values of the boundaries are such that
\ben
 \frac{d-1}{d} < \frac{{\rm Med}(\Delta E\, |\, v, r)}{\overline{\Delta E}} < x \quad \text{with $x$ solution of} \quad \frac{\Gamma(d/2, xd/2)}{\Gamma(d/2)} = \frac{1}{2} \, .
 \label{eq:MedDEBounds}
\een
In practice, this yields $x = \ln(2)$ for $d=2$, and $x = 0.789$ for $d=3$. Eventually, even though the total median value ${\rm Med}(\Delta E)$  cannot be easily computed from  ${\rm Med}(\Delta E \, | \, v, r )$, the properties of the boundaries have to be conserved. Therefore, we have shown that whatever the distribution of initial velocities, the median kinetic energy kick is always equivalent to the average up to an $\mathcal{O}(1)$ pre-factor.

\subsection{The example of a  Maxwellian initial velocity}
\label{app:pdf_ccdf_DeltaE}

In order to illustrate the theoretical development above and to connect with the main text, we now consider that the PDF of initial velocity is a Maxwell-Boltzmann distribution such that
\ben
p_{\vv}(v\, | \, r) = \frac{1}{(2\pi\sigma_{\rm sub}^2(r))^{3/2}}e^{-\frac{v^2}{2\sigma_{\rm sub}^2(r)}}  \, .
\label{eq:MBDistrib}
\een
In order to compute the median of $\Delta E$, we can now compute the exact PDF and CDF. More precisely, similarly to the final velocity in \refeq{finalVelocity}, the PDF for $\Delta E$ is given by 
\ben
p_{\Delta E} (\Delta E \, | \, r)  & = &  \int \dd^3 \vv \,  p_{\vv}(v\, | \, r) \int \dd^d \Delta \vv \,  p_{\Delta \vv}(\Delta v\, | \, r) \delta_{\rm D}\[\Delta E - \frac{(\Delta v)^2}{2} + \vv. \Delta \vv \] \\
 & = &  \int \dd^d \Delta \vv \,  p_{\Delta v}(\Delta v\, | \, r) \frac{2\pi}{\Delta v}\int_{\frac{|2\Delta E - (\Delta v)^2|}{2\Delta v}} \dd v \,  p_{\vv}(v\, | \, r)  \\
  & = &  \int \dd^d \Delta \vv \,  p_{\Delta v}(\Delta v\, | \, r) \frac{1}{\sqrt{2 \pi \sigma_{\rm sub}^2 (\Delta v)^2}}\exp\({\frac{\( \Delta E - \frac{(\Delta  v)^2}{2}  \)^2 }{2\sigma_{\rm sub}^2 |\Delta \vv|^2}} \)\,.
\een
Note that here we integrate over $v$ before integrating over $\Delta v$, for simplicity. Depending on the hypersurface upon which the random walk takes place, we have
\ben
p_{\Delta E}\(\Delta E  \, | \, r\) = 
\begin{cases}
\frac{\exp\(\frac{\Delta E}{2  \sigma_{\rm sub}^2}- \frac{|\Delta E|} {2  \sigma_{\rm sub}^2} \frac{\sqrt{1+s^2}}{s}\)}{4  \sigma_{\rm sub}^2  s \sqrt{1+s^2}}  \quad & {\rm if}  \, \, d = 2 \\[10pt]
 \frac{|\Delta E|e^{\frac{\Delta E}{2\sigma_{\rm sub}^2}}}{4\pi \sigma_{\rm sub}^4 s^2 \sqrt{1+s^2}}  K_1\[\frac{|\Delta E|}{2\sigma_{\rm sub}^2}\frac{\sqrt{1+s^2}}{s}\] \quad & {\rm if}  \, \, d = 3
 \label{eq:pdfDE_vs_d}
\end{cases}
\een
where $s^2 \equiv u^2/(8\sigma_{\rm sub}^2) = \Nc \overline{\delta v^2}/(4d\sigma^2_{\rm sub}) = \overline{\Delta E}/(2d\sigma_{\rm sub}^2)$, and $K_1$ is the modified Bessel function of the second kind of order 1. Now, we focus on the $d=2$ case, and write down the CDF as
\ben
F(<\Delta E \, | \, r) =
\begin{cases}
 \displaystyle 1-\frac{1+\xi}{2\xi}e^{-\frac{\Delta E}{2\sigma^2}(\xi-1)} \quad & {\rm if} \quad \Delta E \ge 0 \\
 \displaystyle \frac{\xi-1}{2\xi}e^{\frac{\Delta E}{2\sigma^2}(1+\xi)} \quad & {\rm else.}
\end{cases}
\een
with $\xi = \sqrt{1+s^2}/s$. The complementary CDF, \ie{} CCDF, introduced in the main text, is denoted $\overline F(> \Delta E) \equiv 1 - F(< \Delta E)$.  Eventually, the energy $\Delta E(q)$ is defined implicitly through the CCDF as the solution of $\overline F(> \Delta E(q)) =q$. Therefore, depending on the value of $q$, the energy $\Delta E (q)$ can be written 
\beq
\begin{split}
 \frac{\Delta E(q)}{2\sigma^2} =   \begin{cases}
\displaystyle\frac{1}{1-\xi}\ln\(\frac{2q\xi}{1+\xi}\)\quad & {\rm if} \quad q < \frac{1+\xi}{2\xi}\\
\displaystyle \frac{1}{1 + \xi}\ln\(\frac{2(1-q)\xi}{\xi -1}\) \quad & {\rm else} \, .
\end{cases}
\end{split}
\eeq
As the average kinetic energy kick is such that $\overline{\Delta E}/(2\sigma_{\rm sub}^2) = 2s^2 = 2(\xi^2-1)$, when $q\le 0.5$, this yields 
\ben
\frac{{\rm Med}(\Delta E)}{\overline{\Delta E}} = \frac{\xi+1}{2}\ln\(\frac{1+\xi}{2q\xi}\)\ge \frac{\xi(1-2q)+1}{2} \ge \frac{1}{2}\,,  
\een
with ${\rm Med}(\Delta E) = \Delta E(0.5)$, and where we used the inequality $\ln(1/x) \ge  1-x$. Eventually, when $q\neq 0.5$ the ratio diverges as $\xi \to \infty$ (corresponding to $s\to 0$). When $q = 0.5$, the ratio is a decreasing function of $\xi$ and
\ben
\frac{{\rm Med}(\Delta E)}{\overline{\Delta E}} \le \lim_{\xi \to 1} \frac{{\rm Med}(\Delta E)}{\overline{\Delta E}} = \ln(2)\, .
\een
which proves the result of \refeq{MedDEBounds}. 

\subsection{Impact of stellar encounters on the phase space distribution function and on the  density profile}
\label{app:ImpactOnProfile}

In this last part, we develop a method to evaluate the new phase-space distribution function (PSDF) after one crossing of the disk assuming that the system is initially isotropic, and that isotropy is preserved. If we know the initial PSDF $f(v, r)$, then trading $p_{\vv}( v\, | \, r)$ for $f$ in \refeq{pvf} allows to define 
\ben
\tilde{f} (v_f, r) =  \frac{1}{4\pi v_f^2}\int \dd^3 \vv  \, f(v, r) p_{v_f}(v_f\, | \, v, r) \, .
\een
We can determine a new profile right after disk crossing (before any relaxation effect that can further change the profile afterward) from the initial one by removing all particles with a final velocity higher than the escape velocity. This amounts to integrate $\tilde{f}$ on $v_f$ between 0 and the escape velocity,
\ben
v_{\rm es}(r) = \sqrt{2|\Phi(r)|}\,,
\een
where $\Phi(r)$ is calculated with the initial profile before any possible reorganization/relaxation\footnote{Note that this approach encompasses a similar method, introduced in \citeref{BerezinskyEtAl2014}, in which case it is possible to properly define a given velocity kick $\delta v$ and $v_f = v+ \delta v$. Then one recovers the same expression by setting
\ben
F_{v_f}(v_{\rm es}(r)\, | \, v, r)  = \Theta\[v_{\rm es}(r) - (v+\delta v)\]\nn\,.
\een
}:
\ben
\tilde{\rho}(r) = \int_{v_f < v_{\rm es}(r)} \dd^3 \vv_f \tilde{f} (v_f, r) = \int \dd^3 \vv  \, f(v, r) F_{v_f}( v_{\rm es}(r) \, | \, v, r) \, .
\een 
The next step is to evaluate the initial PSDF $f$. The common method when the system is assumed spherically symmetric and isotropic is to use the Eddington inversion \cite{Eddington1916a,BinneyEtAl2008}. However, this formalism is plagued with divergences when the initial density profile is truncated at a finite radius. Finding a fully consistent PDF for a NFW profile with a sharp truncation is non-trivial \cite{LacroixEtAl2018}. In practice, one works in positive-definite effective energy space, defining the pseudo-energy $\Ec \equiv \Psi - v^2/2$, with the positive-definite gravitational potential $\Psi = |\Phi|$. For an isotropic system, the PSDF only depends on $\Ec$, and the Eddington inversion provides
\ben
f(\Ec) = \frac{1}{\sqrt{8}\pi^2}\left\{\frac{1}{\sqrt{\Ec}} \left.\der{\rho}{\Psi} \right|_{\Psi = 0}  + \int_0^{\Ec} \frac{\dd \Psi}{\sqrt{\Ec-\Psi}} \der{^2 \rho}{\Psi^2} \right\}\,,
\een
with the relation between $\Psi$ and $\rho$ given by Poisson's equation: $\Delta \Psi = -4\pi G\rho$. When $\Psi = 0$ at finite radius (a possible definition of tidal radius), the derivative of $\rho$ with respect to $\Psi$ does not vanish, and the first term $\propto 1/\sqrt{\Ec}$ diverges and gets nonphysical. This can be fully regularized, as done in \citeref{LacroixEtAl2018}, but it is numerically expensive. In order to keep things simple in the following, we naively remove this term and compute the associated profile as
\ben
\rho(r) = 4\pi \sqrt{2} \int_0^{\Psi(r)} \left\{F(\Ec) \equiv  \int_0^{\Ec} \frac{\dd \Psi'}{\sqrt{\Ec-\Psi'}} \der{^2 \rho}{\Psi'^2} \right\} \sqrt{\Psi(r) - \Ec} \dd \Ec \, .
\een
Two major caveats can be pointed out here. First, the reconstructed density is defined up to a constant that we consider to be 0 here. Moreover, when modifying the PSDF, we actually also modify the density profile and then should also solve for a new gravitational potential in Poisson's equation to get the correct mapping between potential space and real space. The potential does no longer derive from an NFW profile but goes to 0 at the truncation radius, while the impact of stars is evaluated with an exact NFW. Nevertheless, the PSDF in terms of velocity can be evaluated as
\ben
f(v, r) = F\(\Ec = \left\{ \Psi - \frac{v^2}{2} \right\}\).
\label{eq:PSDF_Edd}
\een
In the following, in order to parameterize our uncertainty due to the previously mentioned caveats, we introduce two other simple PSDF. First we consider the Maxwell-Boltzmann distribution \refeq{MBDistrib} used in the main text and we simply set
\ben
f(v, r) = \frac{\rho(r)}{(2\pi\sigma_{\rm sub}^2(r))^{3/2}}e^{-\frac{v^2}{2\sigma_{\rm sub}^2(r)}}  \, .
\label{eq:PSDF_MB_Simp}
\een
This distribution is realistic \cite{ErraniEtAl2021}, but its main downside is that the velocity of particles can be higher than the escape velocity, especially in the outskirts of the structure where the velocity dispersion given by the Jeans equation tends to the gravitational potential. Consequently, close to the truncation radius, the PSDF is not correctly normalized and under-evaluates the density. The second option we investigate is to set a cutoff in velocity space and renormalize,
\ben
f(v, r) = \frac{\rho(r)}{K(r)}\[e^{-\frac{v^2}{2\sigma_{\rm sub}^2(r)}} - e^{-\frac{v_{\rm es}^2(r)}{2\sigma_{\rm sub}^2(r)}} \]\,,
\een
where we set the normalization factor $K(r)$ such that
\ben
\int \dd^3 \vv \,f(v, r) = \rho(r)
\label{eq:PSDF_MB_Trunc}
\een
is ensured. Nevertheless, the main issue with this distribution is that the velocity dispersion is no longer $\sigma_{\rm sub}$ (the posterior velocity dispersion being lower). Unfortunately, trading $\sigma_{\rm sub} \to \sigma \neq \sigma_{\rm sub}$ and trying to recover the correct velocity dispersion does not work either in the outskirts of the halo, as no good value of $\sigma$ can be found. Nonetheless, let us point out that the latter two configurations give a theoretical uncertainty in the outskirts.

In the left panel of \refig{PSDF_New_Profile}, we represent the PSDF with respect to the velocity at a fixed position in a typical subhalo after one disk crossing. We observe, as expected, that the stellar encounters naturally shift the distributions to higher values of the velocity. In the right panel, we show the corresponding new profile density in the top part, and the comparison to the initial profile at the bottom for the different initial PSDFs introduced above. We plotted the result for a subhalo crossing the disk at three different distances from the Galactic center. We can remark that even in the most conservative case of the initial truncated Maxwell Boltzmann distribution -- \refeq{PSDF_MB_Trunc} -- the density decreases toward the outskirts. At $R=$8~kpc and $R=$4.5~kpc, the vertical dash-dotted lines represent the tidal radius obtained from the SL17 recipe, using a typical kinetic energy kick equal to the median. In both cases, they correspond to positions where the new densities are already below $50$\% of the initial one. At $R=$1 kpc the tidal radius is evaluated to be 0 as a sizable fraction of the central particles is ejected (the blue curves being below the $50$\%-threshold in the bottom panel).

We can also remark from the bottom part of the right panel that the central density profile is found to be depleted from its initial value from this simple calculation (minor effect at 8~kpc, while already a 25\%-effect at 4.5~kpc). To determine the true final profile, however, we should also account for relaxation after the shock, which can goes beyond the scope of this paper. We still note, incidentally, that our simple description is in surprisingly good qualitative agreement with simulation results \cite{ErraniEtAl2021,StenDelos2019}.

Overall, this simple analysis is enough to justify the SL17 recipe for the truncation radius, and for the use of the median as a typical energy kick. Further developments are still necessary to better account for changes in the inner density profile, which will be studied elsewhere.

\begin{figure*}[!t]
\centering
\includegraphics[width = 0.49\textwidth]{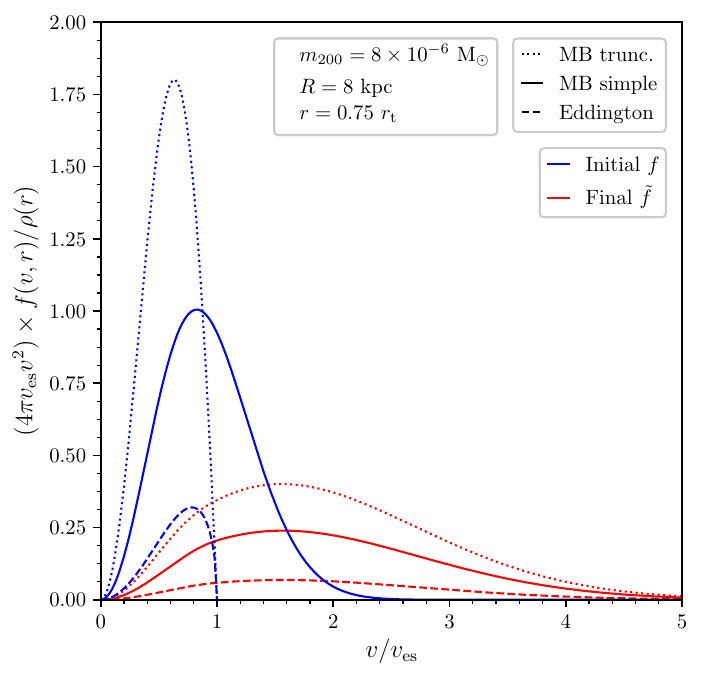}
\includegraphics[width = 0.49\textwidth]{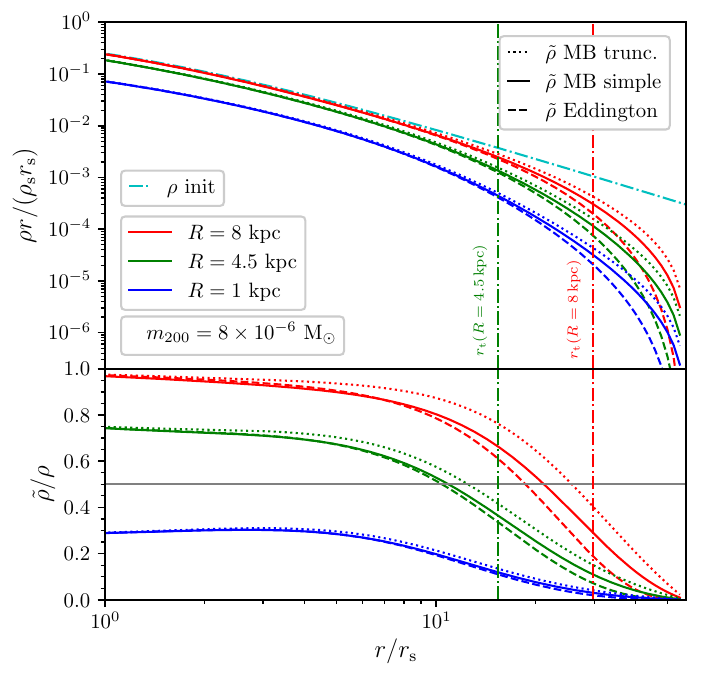}
\caption{\small  {\bf Left panel}. Final PSDFs after energy kick (red curves) in terms of particle speed at a given radius $r$ for a subhalo of mass $m_{\rm 200} = 8\times 10^{-6}$ M$_\odot$ and median concentration crossing the disk at 8~kpc from the center of a MW-like galaxy. The corresponding initial PSDFs are shown as blue curves. The value of $r$ is chosen as 75\% of the tidal radius fixed by the virial radius $r_{200}$. The Eddington-inversion PSDF is given by \refeq{PSDF_Edd}, MB by \refeq{PSDF_MB_Simp}, and truncated-MB by \refeq{PSDF_MB_Trunc}. The normalization was chosen such that the integral between $v=0$ and $v=v_{\rm es}$ of these curves should be 1 to recover the correct value of the density profile at position $r$. The only correctly normalized distribution is MB truncated. The MB PSDF is normalized over the range $v \in [0, \infty [$  and the Eddington PSDF is not normalized at all. The latter two, therefore, under-predict the value of the density. {\bf Right top panel}. The initial profile in cyan and the new profiles computed with the different initial PSDF for a NFW subhalo crossing the disk at three distances from the Galactic center. {\bf Right bottom panel}. Ratio of the new profile to the initial one. The dot-dashed lines correspond to the tidal radii evaluated from the SL17 recipe by choosing a typical kinetic energy kick of every shell at the median value.} 
\label{fig:PSDF_New_Profile}
\end{figure*}

\section{Probability distributions of the total energy kick (stars + disk shocking)}
\label{app:pdf_ccdf_DeltaEtilde}
\begin{figure}[!t]
\centering
\includegraphics[width = 0.5\textwidth]{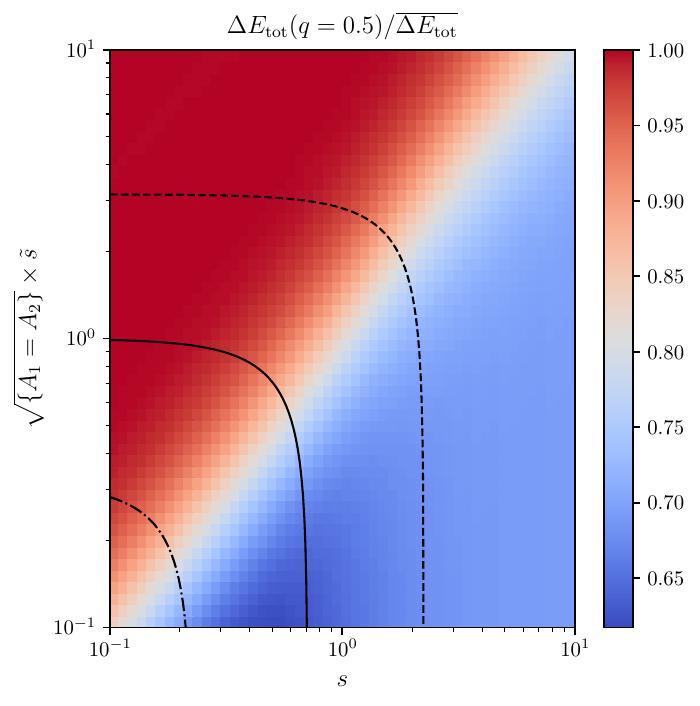}
\caption{\small  $\Etot(q = 0.5)/\overline{\Etot}$ vs the parameters $s$ and $s_{\rm d}$ under the assumption that $A_2 = A_2$ for simplicity.  The black curves represent contours of constant $\overline{\Etot}$: $\overline{\Etot} = 0.2 \sigma^2$ (dash-dotted),  $\overline{\Etot} = 2 \sigma^2$ (solid), $\overline{\Etot} = 20 \sigma^2$ (dashed).}  
\label{fig:DeltaEq05tot}
\end{figure}

As mentioned in the main text, we define a total energy kick as $\Etot = \Delta E + \Delta  E_{\rm d}$, where $\Delta E_{\rm d}$ is the energy kick due to the disk shocking, and $\Delta E$ is the energy kick due to the encounters with stars. In the following, we assume that $\Delta E$ is distributed as in \refeq{pdfDE_vs_d} in the $d=2$ case, and that $\Delta E_{\rm d}$ follows a Gaussian distribution,
\ben
p_{\Delta E_{\rm d}}(\Delta E_{\rm d}) = \frac{1}{\sqrt{2\pi \sigma_{\rm sub}^2 (\Delta \vv_{\rm d})^2 A_2(\eta)}}e^{- \frac{\(\Delta E_{\rm d} - A_1(\eta) \frac{(\Delta \vv_{\rm d})^2}{2}\)^2}{2\sigma_{\rm sub}^2 (\Delta \vv_{\rm d})^2 A_{2}(\eta)} }\,,
\een 
where $\Delta \vv_{\rm d}$ is given  in \refeq{Delta_vv_d}, and the adiabatic correction $A_1$ is introduced in \refeq{Delta_E_d}. We also introduce a new adiabatic corrective factor, $A_2(\eta)$, for the dispersion (according to \citeref{GnedinEtAl1999b}). In the following, we also use the parameter $s_{\rm d} \equiv (\Delta \vv_{\rm d})^2/(4\sigma_{\rm sub}^2)$, similarly to $s = \overline{\Delta E}/(2d\sigma_{\rm sub}^2)$ introduced above. Then, it is possible to define a PDF for $\Etot$. In order to do so, we further introduce two variables,
\ben
\nu_\pm \equiv \frac{1}{2}\(\frac{\sqrt{1+s^2}}{s} \pm 1\)\sqrt{2A_2 s_{\rm d}^2}\,,
\een
and we define a pseudo centered reduced variable corresponding to $\Etot$, of the form
\ben
\eps \equiv \frac{1}{\sqrt{2A_2 s_{\rm d}^2}}\(\frac{\Etot}{2\sigma_{\rm sub}^2} - s_{\rm d}^2 A_1\)\,,
\een
in order to simplify the expressions. The main goal of this analysis is to quantify the asymmetry of the PDF around the average value, so as to evaluate whether considering an average value for $\eps$ is relevant and, if not, what a better choice should be. Therefore, using an affine shift to define this new variable, we do not, a priori, lose in generality. With these two definitions, we can evaluate a PDF for $\eps$ under the form
\beq
\begin{split}
p_\eps(\eps\, |\, \nu_- , \nu_+)  =  \frac{\nu_+ \nu_-}{\nu_+ + \nu_-} e^{-2\eps\nu_- + \nu_-^2}{\rm erfc}(\nu_- - \eps)  + \frac{\nu_+ \nu_-}{\nu_+ + \nu_-} e^{2\eps\nu_+ + \nu_+^2}{\rm erfc}(\nu_+ + \eps) \, .
\end{split}
\eeq
The average value of $\eps$ can be rewritten $\overline{\eps} = (\nu_+ - \nu_-)/(2\nu_+\nu_-)$. Moreover, from this PDF we can also derive a CCDF for $\eps$. This yields the following expression 
\beq
\begin{split}
\overline F_\eps (\eps\, |\, \nu_-, \nu_+)  =  - \frac{\nu_+ \nu_-}{\nu_+ + \nu_-} \frac{e^{2\eps\nu_+ + \nu_+^2}}{2\nu_+}{\rm erfc}(\nu_+ + \eps)   + \frac{\nu_+ \nu_-}{\nu_+ + \nu_-} \frac{e^{-2\eps\nu_- + \nu_-^2}}{2\nu_-}{\rm erfc}(\nu_- - \eps)  +\frac{1}{2}{\rm erfc}(\eps)\, .
\end{split}
\eeq
The main advantage of this parameterization is to only depends on two parameters $\nu_+$ and $\nu_-$, and therefore to be easy to compute numerically. Moreover, it can be used to show that the CCDF evaluated at $\epsilon = \overline \epsilon$ satisfies
\ben
\frac{1}{e} \sim 0.368 < \overline F_\eps (\overline{\eps}\, |\, \nu_+, \nu_-)  < \frac{1}{2} \, .
\een
Considering the average value is a good way to show that the maximal energy gain for a fraction of at least 37\% of the particle. This is understandable as we had, ${\rm Med}(\Etot) \le \overline{\Delta E}$ in the case without disk shocking, and because the addition of $\Delta E_{\rm d}$, which has a symmetric distribution, does not further {\em asymmetrize} the PDF of $\Etot$. Therefore, this justifies entirely that we can evaluate the total energy as roughly being the sum of the averages. In \refig{DeltaEq05tot}, we represent ${\rm Med}(\Delta E)/\overline{\Etot}$ vs. $s$ and $s_{\rm d}$ under the assumption $A_1(\eta) = A_2(\eta)$, for simplicity. When $s_{\rm d} \gg s$, the distribution is symmetrized again with respect to the average, and the ratio is close to $1$. When $s \gg s_{\rm d}$, the symmetry is maximally broken and that yields a ratio of $\sim 0.7$, as discussed in the scenario in which only stellar encounters impact on the subhalos.

\end{widetext}

\end{document}